\documentclass[a4paper]{amsart}
\usepackage[a4paper,margin=25mm]{geometry}
\usepackage{amsmath}
\usepackage{amssymb}
\usepackage{url}

\parskip\smallskipamount

\let\set\mathbb
\def\<#1>{\langle#1\rangle}

\begin{document}

 \author[Manuel Kauers]{Manuel Kauers}
 \address{Manuel Kauers, Institute for Algebra, J. Kepler University Linz, Austria}
 \email{manuel.kauers@jku.at}

 \author[Jakob Moosbauer]{Jakob Moosbauer}
 \address{Jakob Moosbauer, Institute for Algebra, J. Kepler University Linz, Austria}
 \email{jakob.moosbauer@jku.at}
 
 \title{Some New Non-Commutative Matrix~Multiplication Algorithms~of~Size $(n,m,6)$}

 \begin{abstract}
   For various $2\leq n,m\leq6$, we propose some new algorithms for
   multiplying an $n\times m$ matrix with an $m\times6$ matrix
   over a possibly noncommutative coefficient ring.
 \end{abstract}

 \maketitle

 \section{Introduction}

 For given $n,m,p\in\set N$, the standard algorithm for multiplying an $n\times m$-matrix
 with an $m\times p$-matrix requires $nmp$ multiplications in the ground ring, and some
 additions. Strassen~\cite{St:Gein} showed that for $(n,m,p)=(2,2,2)$, it is possible to do the
 job with only seven instead of the usual eight multiplications, and since then, similar
 improvements have been found for other formats $(n,m,p)$. Yet, for most formats, we do
 not know what the minimal required number of multiplications is, and so it is of interest
 to search for matrix multiplication schemes that need fewer multiplications.
  
 In a recent paper~\cite{kauers22i}, we introduced a new technique for searching for
 efficient bilinear multiplication algorithms for matrices of small sizes and
 applied it to all formats $(n,m,p)$ with $2\leq n,m,p\leq 5$. In all instances,
 we either matched the previously smallest known number of required multiplications
 or even found improvements. Our technique is based on investigating random paths
 in a certain graph, see~\cite{Sm:Tbca,ss:ttro,HKS:Nwtm,FBH+:Dfmm} for other search techniques that have
 successfully been applied in the quest for reducing the number of multiplications
 for certain formats. 

 In this short communication, we report on results produced by our technique for matrix
 formats that are slightly larger than those considered in the original paper.
 More precisely, we applied the method to all formats $(n,m,6)$ with $n,m\in\{2,\dots,6\}$.
 This may seem like a minor step forward, but it must be noted that a small increase
 of the matrix size amounts to a substantial increase of the search space, thus
 considerably reducing the chances of finding something new. Indeed, it turned out that
 we are no longer able to match the smallest known number of required multiplications in all cases.
 This may partly be due to inherent limitations of the method and partly be due to limited
 computing resources. For example, the largest case $(n,m,k)=(6,6,6)$ ran for several months
 on a cluster with 1000 cpus but only got down to a scheme with 164 multiplications, while
 the currently best known scheme requires 160~\cite{fastmm}.

 However, for some formats we did manage to reduce the number of multiplications. 

 \section{Matrix Multiplication}

 From an algebraic perspective~\cite{BCS:ACT,landsberg11}, matrix multiplication for a
 given format $(n,m,p)$ and a ground field~$K$ is defined as the tensor
 \[
   \sum_{i=1}^n\sum_{j=1}^m\sum_{k=1}^p a_{i,j}\otimes b_{j,k}\otimes c_{k,i}\in K^{n\times m}\otimes K^{m\times p}\otimes K^{p\times n},
 \]
 where $a_{u,v},b_{u,v},c_{u,v}$ refer to the matrices in $K^{n\times m}, K^{m\times p}, K^{p\times n}$,
 respectively, that have a $1$ at position $(u,v)$, and zeros at all other positions.

 The number of multiplications needed to perform a matrix multiplication is connected to
 the \emph{rank} of the matrix multiplication tensor.
 The rank of a tensor $T\in K^{n\times m}\otimes K^{m\times p}\otimes K^{p\times n}$ is defined
 as the smallest number $r$ such that $T$ can be written as a sum of $r$ tensors of the form
 $M_1\otimes M_2\otimes M_3$.
 For example, the matrix multiplication tensor for $(n,m,k)=(2,2,2)$ can be written as
 \begin{alignat*}1
   &a_{1,1}\otimes b_{1,1}\otimes c_{1,1}\\
   &+a_{1,2}\otimes b_{2,1}\otimes c_{1,1}\\
   &+a_{1,1}\otimes b_{1,2}\otimes c_{2,1}\\
   &+a_{1,2}\otimes b_{2,2}\otimes c_{2,1}\\
   &+a_{2,1}\otimes b_{1,1}\otimes c_{1,2}\\
   &+a_{2,2}\otimes b_{2,1}\otimes c_{1,2}\\
   &+a_{2,1}\otimes b_{1,2}\otimes c_{2,2}\\
   &+a_{2,2}\otimes b_{2,2}\otimes c_{2,2},
 \end{alignat*}
 which amounts to its definition and to the classical matrix multiplication algorithm, but it
 can also be written in the form 
 \begin{alignat*}1
   &(a_{1,1} + a_{2,2}) \otimes (b_{1,1} + b_{2,2}) \otimes (c_{1,1}+c_{2,2})\\
   &+(a_{2,1} + a_{2,2}) \otimes (b_{1,1}) \otimes (c_{1,2}-c_{2,2})\\
   &+(a_{1,1}) \otimes (b_{1,2} - b_{2,2}) \otimes (c_{2,1}+c_{2,2})\\
   &+(a_{2,2}) \otimes (b_{2,1} - b_{1,1}) \otimes (c_{1,1}+c_{1,2})\\
   &+(a_{1,1} + a_{1,2}) \otimes (b_{2,2}) \otimes (c_{2,1}-c_{1,1})\\
   &+(a_{2,1} - a_{1,1}) \otimes (b_{1,1}+ b_{1,2}) \otimes (c_{2,2})\\
   &+(a_{1,2} - a_{2,2}) \otimes (b_{2,1} + b_{2,2}) \otimes (c_{1,1}),
 \end{alignat*}
 which amounts to Strassen's algorithm and shows that the rank for this size is (at most)~7.

 The search for faster matrix multiplication algorithms thus amounts to finding ways to write
 the matrix multiplication tensor as a sum of as few as possible rank-1 tensors. While the
 rank of the tensor is defined as the smallest possible number of summands in such a representation,
 it is convenient to also speak of the rank of a particular representation of the tensor
 (or equivalently, of the rank of a particular matrix multiplication scheme). For example,
 we would say for $(n,m,p)=(2,2,2)$ that the standard algorithm has rank~8 and Strassen's
 algorithm has rank~7.

 It is known~\cite{BCS:ACT} that the rank of a tensor in general depends on the choice of the ground
 field, and there are indications that there may also be such rank mismatches for matrix
 multiplication tensors. For example, Fawzi et al.~\cite{FBH+:Dfmm} recently found an algorithm of rank~47
 for $(n,m,p)=(4,4,4)$ over fields of characteristic~$2$, while the best known algorithm for
 $(n,m,p)=(4,4,4)$ over fields of any other characteristic has rank~49. We found similar mismatches
 for $(n,m,p)=(4,4,5)$ and $(n,m,p)=(5,5,5)$~\cite{kauers22i}.

 \section{Results}

 For all the formats $(n,m,6)$ with $2\leq n,m\leq 6$, here are the smallest ranks that we found
 following random paths in the flip graph starting from the standard algorithm, as in~\cite{kauers22i}.
 The ``naive rank'' is just $6nm$ and refers to the number of multiplications done by the standard
 algorithm. The column ``best rank'' refers to the current record as stated in Sedoglavic's table~\cite{fastmm}.
 A star indicates that the best known scheme given in this table involves fractional coefficients
 and thus only applies under certain restrictions on the characteristic. 

 \begin{center}
 \begin{tabular}{c|c|c|c}
   format & naive & best & our \\[-2pt]
          & rank & rank & rank \\\hline
   $(2,2,6)$ & 24 & 21 & 21 \\
   $(2,3,6)$ & 36 & 30 & 30 \\
   $(2,4,6)$ & 48 & 39 & 39 \\
   $(3,3,6)$ & 54 & 40\rlap{$^\ast$} & 42 \\ 
   $(2,5,6)$ & 60 & 48 & 48 \\
   $(3,4,6)$ & 72 & 56\rlap{$^\ast$} & 56 \\ 
   $(2,6,6)$ & 72 & 57 & 56 \\ 
   $(3,5,6)$ & 90 & 70\rlap{$^\ast$} & 71 \\ 
   $(4,4,6)$ & 96 & 73\rlap{$^\ast$} & 74 \\ 
   $(3,6,6)$ & 108 & 80\rlap{$^\ast$} & 85 \\ 
   $(4,5,6)$ & 120 & 93 & 93 \\
   $(4,6,6)$ & 144 & 105 & 116 \\ 
   $(5,5,6)$ & 150 & 116 & 116 \\
   $(5,6,6)$ & 180 & 137\rlap{$^\ast$} & 144 \\ 
   $(6,6,6)$ & 216 & 160\rlap{$^\ast$} & 164 \\ 
 \end{tabular}
 \end{center}

Remarks:
\begin{itemize}
\item The reduction from 57 to 56 for the format $(2,6,6)$ is the first improvement for this
  format since 1971, when Hopcroft and Kerr~\cite{HK:OMtN} proposed their family of schemes
  for formats of the form $(2,m,p)$.
  To our knowledge, for all $m,p$, the schemes of Hopcroft and Kerr were so far the best known
  for format $(2,m,p)$.
  One of our schemes for the format $(2,6,6)$ is stated below.
\item For the format $(3,3,6)$, the scheme of rank 40 is due to Smirnov~\cite{Sm:Tbca} and has coefficients
  in $\set Z[\frac12]$, so it is not valid for ground fields of characteristic two.
  We do not know whether our bound 42 is best-possible among all schemes that apply to arbitrary
  characteristic. Similarly, our scheme of rank~71 for the format $(3,5,6)$ may be best-possible
  among all schemes that apply to arbitrary characteristic. 
\item For the format $(3,4,6)$, the scheme of rank~56 given in~\cite{fastmm} has fractional
  coefficients. We found schemes of the same rank that only have integer coefficients.
  One of them is given below.
\item For the format $(4,4,6)$, Smirnov~\cite{smirnov23} recently reduced the best-known rank from 75 to~73,
  using a scheme with fractional coefficients. Our scheme is the first that reaches 74 without
  restriction on the characteristic.
\item As the formats get larger, the search becomes more difficult. For the format $(3,6,6)$, 
  we only get down to a scheme of rank~85, but as we reach 42 for the format $(3,3,6)$, it is clear
  that the flip graph for $(3,6,6)$ must also have a path from the standard algorithm to a scheme of
  rank $2\times42=84<85$.
\item For the format $(4,6,6)$, the gap between our rank and the best known rank is particularly large.
  We do not have an explanation for this.
\item For the format $(5,6,6)$, note that combining our rank-56 scheme for $(2,6,6)$ with two copies
  of Smirnov's rank-40 scheme for $(3,3,6)$ yields a scheme of rank $136<137$, albeit with
  fractional coefficients.  We can get a scheme with integer coefficients by using our rank~42
  scheme for $(3,3,6)$ instead.  This results in a scheme of rank-140.  We do not know how good this
  bound is for the rank with no restriction on the characteristic.
\item Finally, for the format $(6,6,6)$, combining a rank-7 scheme for format $(2,2,2)$ with a
  rank-23 scheme for format $(3,3,3)$ gives a scheme of rank 161 without fractional coefficients,
  so our bound 164 is definitely not optimal. In this case however, it is unclear whether we cannot
  reach 161 because there does not exist a path in the flip graph, or if there is a path and we
  are just not able to find it. 
\end{itemize} 

\bibliographystyle{plain}
\bibliography{bib}

\section*{Appendix}

Here is one of our schemes of rank 56 for the format $(2,6,6)$. For typographic reasons, minus signs are
placed above numbers rather than in front of them, e.g., $\bar 2$ means $-2$. 

\allowdisplaybreaks
\footnotesize
\arraycolsep=2pt
\begin{alignat*}3
\begin{pmatrix}
 1 & 0 & 0 & 0 & 0 & \bar1 \\
 \bar1 & 0 & 0 & 0 & 0 & 1 \\
\end{pmatrix}\otimes\begin{pmatrix}
 \bar1 & 0 & 0 & 0 & 0 & 0 \\
 \bar2 & 0 & \bar1 & 0 & \bar1 & 0 \\
 \bar2 & 1 & \bar1 & \bar1 & \bar1 & 1 \\
 \bar2 & 1 & \bar1 & \bar1 & \bar1 & 1 \\
 \bar2 & 0 & \bar1 & \bar1 & \bar2 & 0 \\
 \bar3 & 1 & \bar1 & \bar1 & \bar1 & 1 \\
\end{pmatrix}\otimes\begin{pmatrix}
 0 & 0 \\
 0 & 0 \\
 0 & \bar2 \\
 0 & 0 \\
 0 & 1 \\
 0 & 0 \\
\end{pmatrix}&&+
\begin{pmatrix}
 0 & 0 & 0 & 0 & 0 & 0 \\
 \bar1 & 0 & 0 & 0 & 0 & 0 \\
\end{pmatrix}\otimes\begin{pmatrix}
 0 & 1 & 0 & 0 & 0 & 0 \\
 0 & 1 & 0 & 0 & 0 & 0 \\
 0 & 1 & 0 & 0 & 0 & 0 \\
 0 & 1 & 0 & 0 & 0 & 0 \\
 0 & 0 & 0 & 0 & 0 & 0 \\
 0 & 2 & 0 & 0 & 0 & 0 \\
\end{pmatrix}\otimes\begin{pmatrix}
 0 & 0 \\
 \bar1 & \bar1 \\
 1 & 1 \\
 0 & 0 \\
 0 & 0 \\
 1 & 1 \\
\end{pmatrix}\\+
\begin{pmatrix}
 2 & \bar1 & 1 & 0 & 0 & \bar1 \\
 0 & 0 & 0 & 0 & 0 & 0 \\
\end{pmatrix}\otimes\begin{pmatrix}
 0 & 0 & 0 & 1 & 0 & 0 \\
 0 & 0 & 0 & 1 & 0 & 0 \\
 0 & \bar1 & 0 & 1 & 0 & 0 \\
 0 & \bar1 & 0 & 1 & 0 & 0 \\
 0 & 0 & 0 & 0 & 0 & 0 \\
 0 & 0 & 0 & 1 & 0 & 0 \\
\end{pmatrix}\otimes\begin{pmatrix}
 \bar1 & \bar1 \\
 \bar1 & \bar1 \\
 0 & 0 \\
 0 & 0 \\
 1 & 1 \\
 1 & 1 \\
\end{pmatrix}&&+
\begin{pmatrix}
 \bar1 & 0 & 0 & 0 & 0 & 1 \\
 0 & 0 & 0 & 0 & 0 & 0 \\
\end{pmatrix}\otimes\begin{pmatrix}
 1 & 0 & 0 & 1 & 1 & 0 \\
 2 & 0 & 0 & 2 & 2 & 0 \\
 2 & 0 & 0 & 2 & 2 & 0 \\
 1 & 0 & 0 & 1 & 1 & 0 \\
 2 & 0 & 0 & 2 & 2 & 0 \\
 2 & 0 & 0 & 2 & 2 & 0 \\
\end{pmatrix}\otimes\begin{pmatrix}
 0 & 0 \\
 0 & 0 \\
 \bar2 & \bar2 \\
 0 & 0 \\
 1 & 1 \\
 0 & 0 \\
\end{pmatrix}\\+
\begin{pmatrix}
 1 & 0 & 0 & 0 & 0 & \bar1 \\
 \bar1 & 0 & \bar1 & 0 & 1 & 1 \\
\end{pmatrix}\otimes\begin{pmatrix}
 1 & 0 & 0 & 0 & 0 & 0 \\
 1 & 0 & 0 & 0 & 0 & 0 \\
 1 & 0 & 0 & 0 & 0 & 0 \\
 1 & 0 & 0 & 0 & 0 & 0 \\
 1 & 0 & 0 & 1 & 1 & 0 \\
 2 & 0 & 0 & 0 & 0 & 0 \\
\end{pmatrix}\otimes\begin{pmatrix}
 \bar1 & 0 \\
 0 & 0 \\
 0 & \bar2 \\
 0 & 0 \\
 1 & 1 \\
 0 & 0 \\
\end{pmatrix}&&+
\begin{pmatrix}
 0 & 1 & 0 & 0 & \bar1 & 0 \\
 0 & \bar1 & 1 & \bar1 & 0 & 1 \\
\end{pmatrix}\otimes\begin{pmatrix}
 1 & 0 & 0 & 1 & 0 & 0 \\
 3 & 0 & 1 & 2 & 2 & 0 \\
 3 & 1 & 1 & 2 & 3 & 0 \\
 1 & 1 & 0 & 1 & 1 & 0 \\
 3 & 0 & 1 & 2 & 3 & 0 \\
 3 & 1 & 1 & 2 & 2 & 0 \\
\end{pmatrix}\otimes\begin{pmatrix}
 0 & 0 \\
 0 & 0 \\
 1 & 1 \\
 0 & 0 \\
 0 & 0 \\
 0 & 0 \\
\end{pmatrix}\\+
\begin{pmatrix}
 \bar1 & 0 & \bar1 & 0 & 1 & 1 \\
 1 & 0 & 1 & 0 & \bar1 & \bar1 \\
\end{pmatrix}\otimes\begin{pmatrix}
 0 & 0 & 0 & 0 & 0 & 0 \\
 0 & 0 & 0 & 0 & 0 & 0 \\
 0 & 0 & 0 & 0 & 0 & 0 \\
 0 & 0 & 0 & 0 & 0 & 0 \\
 0 & 0 & 0 & 1 & 1 & 0 \\
 0 & 0 & 0 & 0 & 0 & 0 \\
\end{pmatrix}\otimes\begin{pmatrix}
 \bar1 & 0 \\
 0 & 0 \\
 0 & 0 \\
 0 & 0 \\
 1 & 0 \\
 0 & 0 \\
\end{pmatrix}&&+
\begin{pmatrix}
 0 & 0 & 0 & 0 & 0 & 0 \\
 \bar1 & 1 & \bar1 & 1 & 0 & 0 \\
\end{pmatrix}\otimes\begin{pmatrix}
 \bar1 & 0 & 0 & \bar1 & 0 & 0 \\
 \bar3 & 0 & \bar1 & \bar1 & \bar2 & 0 \\
 \bar3 & 0 & \bar1 & \bar2 & \bar3 & 1 \\
 \bar3 & 0 & \bar1 & \bar2 & \bar3 & 1 \\
 \bar3 & 0 & \bar1 & \bar1 & \bar3 & 0 \\
 \bar3 & 0 & \bar1 & \bar2 & \bar2 & 1 \\
\end{pmatrix}\otimes\begin{pmatrix}
 0 & \bar1 \\
 0 & 0 \\
 0 & 1 \\
 0 & 1 \\
 0 & 0 \\
 0 & 1 \\
\end{pmatrix}\\+
\begin{pmatrix}
 0 & \bar1 & \bar1 & 0 & 1 & 1 \\
 0 & 0 & 0 & 0 & 0 & 0 \\
\end{pmatrix}\otimes\begin{pmatrix}
 0 & 0 & 0 & 1 & 0 & 0 \\
 2 & 0 & 1 & 2 & 2 & 0 \\
 2 & 0 & 1 & 2 & 2 & 0 \\
 0 & 0 & 0 & 1 & 0 & 0 \\
 2 & 0 & 1 & 1 & 2 & 0 \\
 0 & 1 & 0 & 1 & 0 & 1 \\
\end{pmatrix}\otimes\begin{pmatrix}
 0 & 0 \\
 0 & 0 \\
 0 & 0 \\
 0 & 0 \\
 0 & 0 \\
 1 & 0 \\
\end{pmatrix}&&+
\begin{pmatrix}
 0 & 0 & 0 & 0 & 0 & 0 \\
 \bar1 & 1 & 0 & 0 & \bar1 & 0 \\
\end{pmatrix}\otimes\begin{pmatrix}
 0 & 0 & 0 & 0 & 0 & 0 \\
 0 & 0 & 0 & 0 & 0 & 0 \\
 0 & 1 & 0 & 0 & 1 & 0 \\
 0 & 1 & 0 & 0 & 1 & 0 \\
 0 & 1 & 0 & 0 & 1 & 0 \\
 0 & 0 & 0 & 0 & 0 & 0 \\
\end{pmatrix}\otimes\begin{pmatrix}
 0 & 0 \\
 0 & \bar1 \\
 0 & 0 \\
 0 & 0 \\
 0 & 0 \\
 0 & 1 \\
\end{pmatrix}\\+
\begin{pmatrix}
 0 & 0 & 1 & 0 & \bar1 & 0 \\
 0 & 0 & 0 & 0 & 0 & 0 \\
\end{pmatrix}\otimes\begin{pmatrix}
 0 & 0 & 0 & 1 & 0 & 0 \\
 2 & 0 & 1 & 2 & 2 & 0 \\
 1 & 1 & 0 & 1 & 1 & 1 \\
 \bar1 & 1 & \bar1 & 0 & \bar1 & 1 \\
 2 & 0 & 1 & 1 & 2 & 0 \\
 0 & 1 & 0 & 1 & 0 & 1 \\
\end{pmatrix}\otimes\begin{pmatrix}
 1 & 0 \\
 0 & 0 \\
 \bar2 & 0 \\
 0 & 0 \\
 0 & 0 \\
 0 & 0 \\
\end{pmatrix}&&+
\begin{pmatrix}
 \bar1 & 1 & 0 & 0 & 0 & 0 \\
 0 & 0 & 0 & 0 & 0 & 0 \\
\end{pmatrix}\otimes\begin{pmatrix}
 0 & 1 & 0 & 0 & 0 & 1 \\
 0 & 2 & 0 & 0 & 0 & 2 \\
 0 & 2 & 0 & 0 & 0 & 2 \\
 0 & 1 & 0 & 0 & 0 & 1 \\
 0 & 2 & 0 & 0 & 0 & 2 \\
 0 & 2 & 0 & 0 & 0 & 2 \\
\end{pmatrix}\otimes\begin{pmatrix}
 \bar1 & \bar1 \\
 0 & 0 \\
 0 & 0 \\
 0 & 0 \\
 1 & 1 \\
 1 & 1 \\
\end{pmatrix}\\+
\begin{pmatrix}
 \bar1 & 0 & 0 & \bar1 & 0 & 1 \\
 1 & 0 & 0 & 1 & 0 & \bar1 \\
\end{pmatrix}\otimes\begin{pmatrix}
 0 & 0 & 0 & 0 & 0 & 0 \\
 0 & 1 & 0 & 0 & 0 & 0 \\
 0 & 1 & 0 & 0 & 0 & 0 \\
 0 & 1 & 0 & 0 & 0 & 0 \\
 0 & 1 & 0 & 0 & 0 & 0 \\
 0 & 0 & 0 & 0 & 0 & 0 \\
\end{pmatrix}\otimes\begin{pmatrix}
 0 & 0 \\
 \bar1 & 0 \\
 \bar1 & 0 \\
 \bar1 & 0 \\
 1 & 0 \\
 0 & 0 \\
\end{pmatrix}&&+
\begin{pmatrix}
 1 & \bar1 & 0 & 0 & 0 & 0 \\
 0 & 0 & 1 & \bar1 & 0 & 0 \\
\end{pmatrix}\otimes\begin{pmatrix}
 0 & 1 & 0 & 0 & 0 & 1 \\
 0 & 1 & 0 & 1 & 0 & 1 \\
 0 & 2 & 0 & 0 & 0 & 2 \\
 0 & 1 & 0 & 0 & 0 & 1 \\
 0 & 1 & 0 & 1 & 0 & 1 \\
 0 & 2 & 0 & 0 & 0 & 2 \\
\end{pmatrix}\otimes\begin{pmatrix}
 0 & \bar1 \\
 0 & 0 \\
 \bar1 & 0 \\
 \bar1 & 0 \\
 1 & 1 \\
 0 & 1 \\
\end{pmatrix}\\+
\begin{pmatrix}
 0 & 1 & 0 & 0 & \bar1 & 0 \\
 0 & \bar1 & 0 & 0 & 1 & 0 \\
\end{pmatrix}\otimes\begin{pmatrix}
 0 & 0 & 0 & 0 & 0 & 0 \\
 0 & 0 & 0 & 0 & 0 & 0 \\
 0 & 0 & 0 & 0 & 0 & 0 \\
 0 & 0 & 0 & 0 & 0 & 0 \\
 2 & 0 & 1 & 1 & 2 & 0 \\
 0 & 0 & 0 & 0 & 0 & 0 \\
\end{pmatrix}\otimes\begin{pmatrix}
 0 & 0 \\
 0 & 0 \\
 0 & 1 \\
 0 & 0 \\
 0 & 0 \\
 0 & 0 \\
\end{pmatrix}&&+
\begin{pmatrix}
 0 & 0 & 0 & 0 & 0 & 0 \\
 0 & 0 & \bar1 & 1 & 0 & 0 \\
\end{pmatrix}\otimes\begin{pmatrix}
 0 & 0 & 0 & 1 & 0 & 0 \\
 0 & \bar1 & 0 & 2 & 0 & 0 \\
 0 & \bar1 & 0 & 2 & 0 & 0 \\
 0 & 0 & 0 & 1 & 0 & 0 \\
 0 & \bar1 & 0 & 1 & 0 & 0 \\
 0 & 0 & 0 & 2 & 0 & 0 \\
\end{pmatrix}\otimes\begin{pmatrix}
 0 & 0 \\
 0 & 0 \\
 \bar1 & \bar1 \\
 \bar1 & \bar1 \\
 1 & 1 \\
 0 & 0 \\
\end{pmatrix}\\+
\begin{pmatrix}
 0 & \bar1 & 0 & 0 & 1 & 0 \\
 0 & 0 & 0 & 0 & 0 & 0 \\
\end{pmatrix}\otimes\begin{pmatrix}
 0 & 0 & \bar1 & 0 & \bar2 & 1 \\
 2 & 0 & 0 & 1 & 0 & 1 \\
 3 & 1 & 1 & 2 & 3 & 0 \\
 1 & 1 & 0 & 1 & 1 & 0 \\
 3 & 0 & 1 & 2 & 3 & 0 \\
 2 & 1 & 0 & 1 & 0 & 1 \\
\end{pmatrix}\otimes\begin{pmatrix}
 0 & 0 \\
 0 & 0 \\
 1 & 1 \\
 0 & 0 \\
 0 & 0 \\
 0 & 0 \\
\end{pmatrix}&&+
\begin{pmatrix}
 \bar1 & 1 & 0 & 0 & 0 & 0 \\
 1 & \bar1 & 0 & 0 & 0 & 0 \\
\end{pmatrix}\otimes\begin{pmatrix}
 0 & 1 & 0 & 0 & 0 & 1 \\
 1 & 1 & 0 & 1 & 0 & 1 \\
 1 & 2 & 0 & 0 & 0 & 2 \\
 0 & 1 & 0 & 0 & 0 & 1 \\
 1 & 1 & 0 & 1 & 0 & 1 \\
 0 & 2 & 0 & 0 & 0 & 2 \\
\end{pmatrix}\otimes\begin{pmatrix}
 0 & \bar1 \\
 0 & 0 \\
 0 & 0 \\
 0 & 0 \\
 0 & 1 \\
 0 & 1 \\
\end{pmatrix}\\+
\begin{pmatrix}
 \bar1 & 0 & 0 & 0 & 0 & 1 \\
 0 & 1 & \bar1 & 1 & 0 & \bar1 \\
\end{pmatrix}\otimes\begin{pmatrix}
 \bar1 & 0 & 0 & 0 & 0 & 0 \\
 \bar2 & 0 & \bar1 & 0 & \bar1 & 0 \\
 \bar2 & 0 & \bar1 & \bar1 & \bar2 & 1 \\
 \bar2 & 0 & \bar1 & \bar1 & \bar2 & 1 \\
 \bar2 & 0 & \bar1 & \bar1 & \bar2 & 0 \\
 \bar3 & 0 & \bar1 & \bar1 & \bar2 & 1 \\
\end{pmatrix}\otimes\begin{pmatrix}
 0 & 0 \\
 0 & 0 \\
 \bar1 & \bar2 \\
 0 & 0 \\
 0 & 1 \\
 0 & 0 \\
\end{pmatrix}&&+
\begin{pmatrix}
 0 & \bar1 & 0 & 0 & 1 & 0 \\
 0 & 1 & \bar1 & 0 & 0 & 0 \\
\end{pmatrix}\otimes\begin{pmatrix}
 0 & 0 & 0 & 1 & 0 & 0 \\
 2 & 0 & 1 & 2 & 2 & 0 \\
 1 & 1 & 0 & 1 & 1 & 1 \\
 \bar1 & 1 & \bar1 & 0 & \bar1 & 1 \\
 3 & 0 & 1 & 2 & 3 & 0 \\
 0 & 1 & 0 & 1 & 0 & 1 \\
\end{pmatrix}\otimes\begin{pmatrix}
 1 & 0 \\
 0 & 0 \\
 \bar1 & 1 \\
 0 & 0 \\
 0 & 0 \\
 0 & 0 \\
\end{pmatrix}\\+
\begin{pmatrix}
 2 & \bar1 & 1 & 0 & 0 & \bar1 \\
 \bar1 & 0 & \bar1 & 0 & 1 & 1 \\
\end{pmatrix}\otimes\begin{pmatrix}
 0 & 0 & 0 & 0 & 0 & 0 \\
 0 & 0 & 0 & 0 & 0 & 0 \\
 0 & 1 & 0 & 0 & 1 & 0 \\
 0 & 1 & 0 & 0 & 1 & 0 \\
 0 & 0 & 0 & 1 & 1 & 0 \\
 0 & 0 & 0 & 0 & 0 & 0 \\
\end{pmatrix}\otimes\begin{pmatrix}
 \bar1 & \bar1 \\
 0 & \bar1 \\
 0 & 0 \\
 0 & 0 \\
 1 & 1 \\
 0 & 1 \\
\end{pmatrix}&&+
\begin{pmatrix}
 \bar1 & 1 & 1 & \bar1 & \bar1 & 0 \\
 1 & 0 & 0 & 1 & 0 & \bar1 \\
\end{pmatrix}\otimes\begin{pmatrix}
 0 & 1 & 0 & \bar1 & 0 & 1 \\
 0 & 1 & 0 & \bar1 & 0 & 1 \\
 0 & 1 & 0 & \bar1 & 0 & 1 \\
 0 & 1 & 0 & \bar1 & 0 & 1 \\
 0 & 1 & 0 & 0 & 0 & 1 \\
 0 & 1 & 0 & \bar1 & 0 & 1 \\
\end{pmatrix}\otimes\begin{pmatrix}
 0 & 0 \\
 \bar1 & \bar1 \\
 \bar1 & \bar1 \\
 \bar1 & \bar1 \\
 1 & 1 \\
 0 & 1 \\
\end{pmatrix}\\+
\begin{pmatrix}
 2 & 0 & 0 & 0 & 0 & \bar1 \\
 0 & 0 & 0 & 0 & 0 & 0 \\
\end{pmatrix}\otimes\begin{pmatrix}
 0 & 0 & 0 & 0 & 1 & 0 \\
 0 & 0 & 0 & 0 & 1 & 0 \\
 0 & 0 & 0 & 0 & 0 & 0 \\
 0 & 0 & 0 & 0 & 0 & 0 \\
 0 & 0 & 0 & 0 & 0 & 0 \\
 0 & 0 & 0 & 0 & 1 & 0 \\
\end{pmatrix}\otimes\begin{pmatrix}
 0 & 0 \\
 \bar1 & 0 \\
 \bar1 & 0 \\
 0 & 0 \\
 1 & 0 \\
 1 & 0 \\
\end{pmatrix}&&+
\begin{pmatrix}
 1 & \bar1 & 0 & 0 & 0 & 0 \\
 \bar1 & 0 & \bar1 & 0 & 1 & 1 \\
\end{pmatrix}\otimes\begin{pmatrix}
 0 & 1 & 0 & 0 & 0 & 1 \\
 1 & 2 & 0 & 0 & 0 & 2 \\
 1 & 2 & 0 & 0 & 0 & 2 \\
 0 & 1 & 0 & 0 & 0 & 1 \\
 1 & 2 & 0 & 0 & 0 & 2 \\
 0 & 2 & 0 & 0 & 0 & 2 \\
\end{pmatrix}\otimes\begin{pmatrix}
 \bar1 & \bar1 \\
 0 & 0 \\
 0 & 0 \\
 0 & 0 \\
 1 & 1 \\
 0 & 1 \\
\end{pmatrix}\\+
\begin{pmatrix}
 0 & 0 & 1 & 0 & 0 & \bar1 \\
 0 & 0 & \bar1 & 0 & 0 & 1 \\
\end{pmatrix}\otimes\begin{pmatrix}
 0 & 0 & 0 & 0 & 0 & 0 \\
 2 & 0 & 1 & 1 & 2 & 0 \\
 2 & 0 & 1 & 1 & 2 & 0 \\
 0 & 0 & 0 & 0 & 0 & 0 \\
 2 & 0 & 1 & 1 & 2 & 0 \\
 0 & 0 & 0 & 0 & 0 & 0 \\
\end{pmatrix}\otimes\begin{pmatrix}
 0 & 0 \\
 0 & 0 \\
 1 & 0 \\
 0 & 0 \\
 0 & 0 \\
 1 & 0 \\
\end{pmatrix}&&+
\begin{pmatrix}
 0 & 1 & \bar1 & 0 & 0 & 0 \\
 0 & \bar1 & 1 & 0 & 0 & 0 \\
\end{pmatrix}\otimes\begin{pmatrix}
 \bar1 & 1 & 0 & \bar2 & 0 & 1 \\
 \bar3 & 2 & \bar1 & \bar3 & \bar2 & 1 \\
 \bar1 & 2 & 0 & \bar2 & \bar1 & 1 \\
 0 & 1 & 1 & \bar1 & 1 & 0 \\
 \bar3 & 2 & \bar1 & \bar2 & \bar3 & 1 \\
 \bar1 & 1 & 0 & \bar2 & 0 & 1 \\
\end{pmatrix}\otimes\begin{pmatrix}
 \bar1 & 0 \\
 0 & 0 \\
 1 & 0 \\
 0 & 0 \\
 0 & 0 \\
 0 & 0 \\
\end{pmatrix}\\+
\begin{pmatrix}
 0 & \bar1 & 1 & \bar1 & 0 & 1 \\
 0 & 1 & \bar1 & 1 & 0 & \bar1 \\
\end{pmatrix}\otimes\begin{pmatrix}
 0 & 0 & 0 & 1 & 0 & 0 \\
 1 & 0 & 0 & 2 & 1 & 0 \\
 1 & 1 & 0 & 1 & 1 & 1 \\
 \bar1 & 1 & \bar1 & 0 & \bar1 & 1 \\
 1 & 0 & 0 & 1 & 1 & 0 \\
 0 & 1 & 0 & 1 & 0 & 1 \\
\end{pmatrix}\otimes\begin{pmatrix}
 0 & 0 \\
 0 & 0 \\
 1 & 0 \\
 0 & 0 \\
 0 & 0 \\
 0 & 0 \\
\end{pmatrix}&&+
\begin{pmatrix}
 \bar1 & 0 & 0 & 0 & 0 & 1 \\
 \bar1 & 0 & 0 & 0 & 0 & 0 \\
\end{pmatrix}\otimes\begin{pmatrix}
 0 & 0 & 0 & 0 & 1 & 0 \\
 0 & 0 & 0 & 0 & 1 & 0 \\
 0 & \bar1 & 0 & 0 & 0 & 0 \\
 0 & \bar1 & 0 & 0 & 0 & 0 \\
 0 & 0 & 0 & 0 & 0 & 0 \\
 0 & \bar1 & 0 & 0 & 1 & 0 \\
\end{pmatrix}\otimes\begin{pmatrix}
 0 & 0 \\
 \bar1 & 0 \\
 1 & 2 \\
 0 & 0 \\
 0 & \bar1 \\
 1 & 0 \\
\end{pmatrix}\\+
\begin{pmatrix}
 0 & 1 & 0 & 0 & \bar1 & 0 \\
 0 & 0 & 1 & 0 & 0 & \bar1 \\
\end{pmatrix}\otimes\begin{pmatrix}
 0 & 1 & 0 & 0 & 0 & 1 \\
 2 & 1 & 1 & 1 & 2 & 1 \\
 2 & 1 & 1 & 1 & 2 & 1 \\
 0 & 1 & 0 & 0 & 0 & 1 \\
 2 & 0 & 1 & 1 & 2 & 0 \\
 0 & 2 & 0 & 0 & 0 & 2 \\
\end{pmatrix}\otimes\begin{pmatrix}
 0 & 0 \\
 0 & 0 \\
 1 & 1 \\
 0 & 0 \\
 0 & 0 \\
 1 & 0 \\
\end{pmatrix}&&+
\begin{pmatrix}
 0 & 0 & 0 & 0 & 0 & 0 \\
 2 & 0 & 0 & 0 & 0 & \bar1 \\
\end{pmatrix}\otimes\begin{pmatrix}
 1 & 0 & 0 & 1 & 0 & 0 \\
 1 & 0 & 0 & 1 & 0 & 0 \\
 1 & 1 & 0 & 1 & 1 & 0 \\
 1 & 1 & 0 & 1 & 1 & 0 \\
 1 & 0 & 0 & 1 & 1 & 0 \\
 1 & 1 & 0 & 1 & 0 & 0 \\
\end{pmatrix}\otimes\begin{pmatrix}
 0 & 0 \\
 0 & \bar1 \\
 0 & \bar1 \\
 0 & 0 \\
 0 & 1 \\
 0 & 1 \\
\end{pmatrix}\\+
\begin{pmatrix}
 1 & 0 & 0 & 0 & 0 & \bar1 \\
 0 & 0 & 1 & \bar1 & 0 & 0 \\
\end{pmatrix}\otimes\begin{pmatrix}
 0 & 0 & 0 & 1 & 0 & 0 \\
 1 & 0 & 0 & 2 & 1 & 0 \\
 1 & 0 & 0 & 2 & 1 & 0 \\
 0 & 0 & 0 & 1 & 0 & 0 \\
 1 & 0 & 0 & 1 & 1 & 0 \\
 0 & 0 & 0 & 2 & 0 & 0 \\
\end{pmatrix}\otimes\begin{pmatrix}
 0 & 0 \\
 0 & 0 \\
 \bar1 & \bar2 \\
 \bar1 & 0 \\
 1 & 1 \\
 0 & 0 \\
\end{pmatrix}&&+
\begin{pmatrix}
 1 & \bar1 & 1 & \bar1 & 0 & 0 \\
 0 & 0 & 0 & 0 & 0 & 0 \\
\end{pmatrix}\otimes\begin{pmatrix}
 0 & 1 & 0 & 0 & 0 & 1 \\
 1 & 2 & 0 & 1 & 1 & 1 \\
 1 & 3 & 0 & 0 & 1 & 2 \\
 0 & 2 & 0 & 0 & 1 & 1 \\
 1 & 2 & 0 & 1 & 1 & 1 \\
 0 & 2 & 0 & 0 & 0 & 2 \\
\end{pmatrix}\otimes\begin{pmatrix}
 \bar1 & 0 \\
 0 & 0 \\
 2 & 0 \\
 1 & 0 \\
 \bar1 & 0 \\
 1 & 0 \\
\end{pmatrix}\\+
\begin{pmatrix}
 \bar1 & 0 & 0 & 1 & 0 & 0 \\
 0 & 1 & \bar1 & 0 & 0 & 0 \\
\end{pmatrix}\otimes\begin{pmatrix}
 0 & \bar1 & 0 & 1 & 0 & \bar1 \\
 0 & \bar2 & 0 & 1 & 0 & \bar1 \\
 0 & \bar3 & 0 & 1 & 0 & \bar2 \\
 1 & \bar2 & 0 & 1 & 0 & \bar1 \\
 0 & \bar2 & 0 & 0 & 0 & \bar1 \\
 0 & \bar2 & 0 & 1 & 0 & \bar2 \\
\end{pmatrix}\otimes\begin{pmatrix}
 1 & 0 \\
 0 & 0 \\
 \bar1 & 0 \\
 0 & 1 \\
 0 & 0 \\
 0 & 1 \\
\end{pmatrix}&&+
\begin{pmatrix}
 2 & \bar1 & 1 & 0 & 0 & \bar1 \\
 \bar2 & 1 & \bar1 & 0 & 0 & 1 \\
\end{pmatrix}\otimes\begin{pmatrix}
 1 & 0 & 0 & 1 & 0 & 0 \\
 1 & \bar1 & 0 & 1 & 0 & 0 \\
 1 & \bar1 & 0 & 1 & 0 & 0 \\
 1 & \bar1 & 0 & 1 & 0 & 0 \\
 1 & \bar1 & 0 & 0 & 0 & 0 \\
 1 & 0 & 0 & 1 & 0 & 0 \\
\end{pmatrix}\otimes\begin{pmatrix}
 0 & \bar1 \\
 0 & \bar1 \\
 0 & 0 \\
 0 & 0 \\
 0 & 1 \\
 0 & 1 \\
\end{pmatrix}\\+
\begin{pmatrix}
 1 & \bar1 & \bar1 & 1 & 1 & 0 \\
 \bar1 & 1 & 1 & \bar1 & \bar1 & 0 \\
\end{pmatrix}\otimes\begin{pmatrix}
 0 & 0 & 0 & 0 & 0 & 0 \\
 0 & 0 & 0 & 0 & 0 & 0 \\
 0 & 0 & 0 & 0 & 0 & 0 \\
 0 & 0 & 0 & 0 & 0 & 0 \\
 0 & 0 & 0 & 1 & 0 & 0 \\
 0 & 0 & 0 & 0 & 0 & 0 \\
\end{pmatrix}\otimes\begin{pmatrix}
 0 & 0 \\
 0 & \bar1 \\
 0 & \bar1 \\
 0 & \bar1 \\
 0 & 1 \\
 0 & 1 \\
\end{pmatrix}&&+
\begin{pmatrix}
 0 & 0 & 0 & 0 & 0 & 0 \\
 1 & 0 & 1 & \bar1 & 0 & \bar1 \\
\end{pmatrix}\otimes\begin{pmatrix}
 0 & 0 & 0 & 0 & 0 & 0 \\
 0 & 0 & 0 & 0 & 0 & 0 \\
 0 & 1 & 0 & \bar1 & 0 & 1 \\
 0 & 1 & 0 & \bar1 & 0 & 1 \\
 0 & 0 & 0 & 0 & 0 & 0 \\
 0 & 1 & 0 & \bar1 & 0 & 1 \\
\end{pmatrix}\otimes\begin{pmatrix}
 0 & 0 \\
 0 & 0 \\
 0 & \bar1 \\
 0 & 1 \\
 0 & 0 \\
 0 & 0 \\
\end{pmatrix}\\+
\begin{pmatrix}
 0 & \bar1 & 0 & 0 & 1 & 0 \\
 1 & 0 & 0 & 0 & 0 & 0 \\
\end{pmatrix}\otimes\begin{pmatrix}
 1 & 1 & 1 & 0 & 2 & 0 \\
 1 & 1 & 1 & 0 & 2 & 0 \\
 1 & 0 & 1 & 0 & 1 & 0 \\
 1 & 0 & 1 & 0 & 1 & 0 \\
 1 & 0 & 1 & 0 & 1 & 0 \\
 1 & 1 & 1 & 0 & 2 & 0 \\
\end{pmatrix}\otimes\begin{pmatrix}
 0 & 0 \\
 \bar1 & 0 \\
 1 & 1 \\
 0 & 0 \\
 0 & 0 \\
 1 & 0 \\
\end{pmatrix}&&+
\begin{pmatrix}
 0 & 0 & 0 & 0 & 0 & 0 \\
 0 & 0 & \bar1 & 0 & 0 & 1 \\
\end{pmatrix}\otimes\begin{pmatrix}
 0 & 1 & 0 & \bar1 & 0 & 1 \\
 0 & 1 & 0 & \bar1 & 0 & 1 \\
 0 & 0 & 0 & 0 & 0 & 0 \\
 0 & 0 & 0 & 0 & 0 & 0 \\
 0 & 0 & 0 & 0 & 0 & 0 \\
 0 & 1 & 0 & \bar1 & 0 & 1 \\
\end{pmatrix}\otimes\begin{pmatrix}
 0 & 0 \\
 0 & 0 \\
 1 & 1 \\
 0 & 0 \\
 0 & 0 \\
 1 & 1 \\
\end{pmatrix}\\+
\begin{pmatrix}
 0 & 0 & 0 & 0 & 0 & 0 \\
 1 & 0 & 1 & 0 & \bar1 & \bar1 \\
\end{pmatrix}\otimes\begin{pmatrix}
 1 & 1 & 0 & 0 & 0 & 1 \\
 2 & 2 & 0 & 0 & 0 & 2 \\
 2 & 3 & 0 & 0 & 1 & 2 \\
 1 & 2 & 0 & 0 & 1 & 1 \\
 2 & 2 & 0 & 1 & 1 & 2 \\
 2 & 2 & 0 & 0 & 0 & 2 \\
\end{pmatrix}\otimes\begin{pmatrix}
 \bar1 & \bar1 \\
 0 & 0 \\
 0 & 0 \\
 0 & 0 \\
 1 & 1 \\
 0 & 0 \\
\end{pmatrix}&&+
\begin{pmatrix}
 \bar1 & 1 & \bar1 & 1 & 0 & 0 \\
 0 & 0 & 0 & 0 & 0 & 0 \\
\end{pmatrix}\otimes\begin{pmatrix}
 0 & 0 & 0 & 1 & 0 & 0 \\
 1 & 0 & 0 & 2 & 1 & 0 \\
 1 & 0 & 0 & 1 & 1 & 0 \\
 0 & 0 & 0 & 1 & 1 & 0 \\
 1 & 0 & 0 & 1 & 1 & 0 \\
 0 & 0 & 0 & 1 & 0 & 0 \\
\end{pmatrix}\otimes\begin{pmatrix}
 \bar1 & 0 \\
 0 & 0 \\
 1 & 0 \\
 1 & 0 \\
 0 & 0 \\
 1 & 0 \\
\end{pmatrix}\\+
\begin{pmatrix}
 0 & 0 & 0 & 0 & 0 & 0 \\
 0 & 1 & \bar1 & 0 & 0 & 0 \\
\end{pmatrix}\otimes\begin{pmatrix}
 0 & 0 & 0 & 1 & 0 & 0 \\
 0 & 0 & 0 & 1 & 0 & 0 \\
 1 & 0 & 0 & 1 & 1 & 0 \\
 1 & 0 & 0 & 1 & 1 & 0 \\
 1 & 0 & 0 & 1 & 1 & 0 \\
 0 & 0 & 0 & 1 & 0 & 0 \\
\end{pmatrix}\otimes\begin{pmatrix}
 \bar1 & \bar1 \\
 0 & 0 \\
 1 & 1 \\
 0 & 0 \\
 0 & 0 \\
 0 & 0 \\
\end{pmatrix}&&+
\begin{pmatrix}
 \bar1 & 0 & 0 & 1 & 0 & 0 \\
 1 & 0 & 0 & \bar1 & 0 & 0 \\
\end{pmatrix}\otimes\begin{pmatrix}
 0 & 1 & 0 & \bar1 & 0 & 1 \\
 0 & 2 & 0 & \bar1 & 0 & 1 \\
 0 & 2 & 0 & \bar1 & 0 & 1 \\
 \bar1 & 1 & 0 & \bar1 & 0 & 0 \\
 0 & 2 & 0 & 0 & 0 & 1 \\
 0 & 1 & 0 & \bar1 & 0 & 1 \\
\end{pmatrix}\otimes\begin{pmatrix}
 0 & 0 \\
 0 & 0 \\
 0 & 0 \\
 0 & 1 \\
 0 & 0 \\
 0 & 1 \\
\end{pmatrix}\\+
\begin{pmatrix}
 0 & 0 & 0 & 0 & 0 & 0 \\
 \bar1 & 1 & \bar1 & 1 & 0 & 0 \\
\end{pmatrix}\otimes\begin{pmatrix}
 1 & 1 & 0 & 0 & 0 & 1 \\
 3 & 2 & 1 & 1 & 2 & 1 \\
 3 & 3 & 1 & 1 & 3 & 1 \\
 2 & 2 & 1 & 1 & 3 & 0 \\
 3 & 2 & 1 & 2 & 3 & 1 \\
 3 & 2 & 1 & 1 & 2 & 1 \\
\end{pmatrix}\otimes\begin{pmatrix}
 0 & \bar1 \\
 0 & 0 \\
 0 & 0 \\
 0 & 1 \\
 0 & 1 \\
 0 & 1 \\
\end{pmatrix}&&+
\begin{pmatrix}
 0 & 0 & 0 & 0 & 0 & 0 \\
 0 & 0 & 1 & 0 & \bar1 & 0 \\
\end{pmatrix}\otimes\begin{pmatrix}
 1 & 0 & 0 & 0 & 0 & 0 \\
 1 & 0 & 0 & 0 & 0 & 0 \\
 1 & 0 & 0 & 0 & 0 & 0 \\
 1 & 0 & 0 & 0 & 0 & 0 \\
 0 & 0 & 0 & 0 & 0 & 0 \\
 2 & 0 & 0 & 0 & 0 & 0 \\
\end{pmatrix}\otimes\begin{pmatrix}
 0 & 1 \\
 0 & 0 \\
 0 & \bar2 \\
 0 & 0 \\
 0 & 0 \\
 0 & 0 \\
\end{pmatrix}\\+
\begin{pmatrix}
 1 & 0 & 0 & \bar1 & 0 & 0 \\
 0 & 0 & 0 & 0 & 0 & 0 \\
\end{pmatrix}\otimes\begin{pmatrix}
 0 & 0 & 0 & 1 & 0 & 0 \\
 0 & 0 & 0 & 1 & 0 & 0 \\
 0 & 0 & 0 & 0 & 0 & 0 \\
 0 & 0 & 0 & 0 & 0 & 0 \\
 0 & 0 & 0 & 0 & 0 & 0 \\
 0 & 0 & 0 & 1 & 0 & 0 \\
\end{pmatrix}\otimes\begin{pmatrix}
 0 & 0 \\
 0 & 0 \\
 0 & 0 \\
 1 & 1 \\
 0 & 0 \\
 1 & 1 \\
\end{pmatrix}&&+
\begin{pmatrix}
 0 & 0 & 0 & 0 & 0 & 0 \\
 \bar1 & 0 & 0 & \bar1 & 0 & 1 \\
\end{pmatrix}\otimes\begin{pmatrix}
 0 & 1 & 0 & 0 & 0 & 1 \\
 0 & 1 & 0 & 0 & 0 & 1 \\
 0 & 1 & 0 & 0 & 0 & 1 \\
 0 & 1 & 0 & 0 & 0 & 1 \\
 0 & 1 & 0 & 0 & 0 & 1 \\
 0 & 1 & 0 & 0 & 0 & 1 \\
\end{pmatrix}\otimes\begin{pmatrix}
 0 & 0 \\
 \bar1 & \bar1 \\
 \bar1 & \bar1 \\
 \bar1 & \bar1 \\
 1 & 1 \\
 0 & 0 \\
\end{pmatrix}\\+
\begin{pmatrix}
 \bar1 & 1 & 1 & \bar1 & \bar1 & 0 \\
 0 & 0 & \bar1 & 1 & 0 & 0 \\
\end{pmatrix}\otimes\begin{pmatrix}
 0 & 0 & 0 & 0 & 0 & 0 \\
 0 & 1 & 0 & 0 & 0 & 0 \\
 0 & 1 & 0 & 0 & 0 & 0 \\
 0 & 0 & 0 & 0 & 0 & 0 \\
 0 & 1 & 0 & 1 & 0 & 0 \\
 0 & 0 & 0 & 0 & 0 & 0 \\
\end{pmatrix}\otimes\begin{pmatrix}
 0 & 0 \\
 0 & \bar1 \\
 \bar1 & \bar1 \\
 \bar1 & \bar1 \\
 1 & 1 \\
 0 & 1 \\
\end{pmatrix}&&+
\begin{pmatrix}
 \bar1 & 1 & 0 & 0 & \bar1 & 0 \\
 0 & 0 & 0 & 0 & 0 & 0 \\
\end{pmatrix}\otimes\begin{pmatrix}
 1 & 1 & 1 & 0 & 2 & 0 \\
 1 & 2 & 1 & 0 & 2 & 0 \\
 1 & 1 & 1 & 0 & 1 & 0 \\
 1 & 0 & 1 & 0 & 1 & 0 \\
 1 & 2 & 1 & 0 & 1 & 0 \\
 1 & 1 & 1 & 0 & 2 & 0 \\
\end{pmatrix}\otimes\begin{pmatrix}
 0 & 0 \\
 \bar1 & 0 \\
 0 & 0 \\
 0 & 0 \\
 0 & 0 \\
 1 & 0 \\
\end{pmatrix}\\+
\begin{pmatrix}
 1 & \bar1 & \bar1 & 1 & 1 & 0 \\
 0 & 0 & 0 & 0 & 0 & 0 \\
\end{pmatrix}\otimes\begin{pmatrix}
 0 & 1 & 0 & \bar1 & 0 & 1 \\
 0 & 2 & 0 & \bar1 & 0 & 1 \\
 0 & 2 & 0 & \bar1 & 0 & 1 \\
 0 & 1 & 0 & \bar1 & 0 & 1 \\
 0 & 2 & 0 & 0 & 0 & 1 \\
 0 & 1 & 0 & \bar1 & 0 & 1 \\
\end{pmatrix}\otimes\begin{pmatrix}
 0 & 0 \\
 \bar1 & \bar1 \\
 \bar1 & \bar1 \\
 \bar1 & \bar1 \\
 1 & 1 \\
 1 & 1 \\
\end{pmatrix}&&+
\begin{pmatrix}
 0 & 0 & 0 & 0 & 0 & 0 \\
 0 & 1 & 1 & 0 & \bar1 & \bar1 \\
\end{pmatrix}\otimes\begin{pmatrix}
 0 & 1 & 0 & 0 & 0 & 1 \\
 1 & 2 & 0 & 0 & 0 & 2 \\
 1 & 2 & 0 & 0 & 0 & 2 \\
 0 & 1 & 0 & 0 & 0 & 1 \\
 1 & 1 & 0 & 0 & 0 & 1 \\
 0 & 2 & 0 & 0 & 0 & 2 \\
\end{pmatrix}\otimes\begin{pmatrix}
 0 & 0 \\
 0 & 0 \\
 0 & 0 \\
 0 & 0 \\
 0 & 0 \\
 0 & 1 \\
\end{pmatrix}\\+
\begin{pmatrix}
 \bar2 & 1 & \bar1 & 0 & 0 & 1 \\
 1 & 0 & 0 & 1 & 0 & \bar1 \\
\end{pmatrix}\otimes\begin{pmatrix}
 0 & 0 & 0 & 1 & 0 & 0 \\
 0 & \bar1 & 0 & 1 & 0 & 0 \\
 0 & \bar1 & 0 & 1 & 0 & 0 \\
 0 & \bar1 & 0 & 1 & 0 & 0 \\
 0 & \bar1 & 0 & 0 & 0 & 0 \\
 0 & 0 & 0 & 1 & 0 & 0 \\
\end{pmatrix}\otimes\begin{pmatrix}
 0 & \bar1 \\
 \bar1 & \bar1 \\
 \bar1 & 0 \\
 \bar1 & 0 \\
 1 & 1 \\
 0 & 1 \\
\end{pmatrix}&&+
\begin{pmatrix}
 0 & 0 & 1 & \bar1 & 0 & 0 \\
 0 & 0 & \bar1 & 1 & 0 & 0 \\
\end{pmatrix}\otimes\begin{pmatrix}
 0 & 1 & 0 & 0 & 0 & 1 \\
 1 & 1 & 0 & 1 & 1 & 1 \\
 1 & 2 & 0 & 0 & 1 & 2 \\
 0 & 1 & 0 & 0 & 0 & 1 \\
 1 & 1 & 0 & 0 & 1 & 1 \\
 0 & 2 & 0 & 0 & 0 & 2 \\
\end{pmatrix}\otimes\begin{pmatrix}
 0 & 0 \\
 0 & 0 \\
 \bar1 & 0 \\
 \bar1 & 0 \\
 1 & 0 \\
 0 & 0 \\
\end{pmatrix}\\+
\begin{pmatrix}
 \bar2 & 1 & \bar1 & 0 & 0 & 1 \\
 0 & \bar1 & 1 & 0 & 0 & 0 \\
\end{pmatrix}\otimes\begin{pmatrix}
 1 & 0 & 0 & 1 & 0 & 0 \\
 1 & 0 & 0 & 1 & 0 & 0 \\
 1 & 0 & 0 & 1 & 1 & 0 \\
 1 & 0 & 0 & 1 & 1 & 0 \\
 1 & 0 & 0 & 1 & 1 & 0 \\
 1 & 0 & 0 & 1 & 0 & 0 \\
\end{pmatrix}\otimes\begin{pmatrix}
 \bar1 & \bar1 \\
 0 & \bar1 \\
 1 & 0 \\
 0 & 0 \\
 0 & 1 \\
 0 & 1 \\
\end{pmatrix}&&+
\begin{pmatrix}
 1 & 0 & 0 & 0 & 0 & 0 \\
 \bar1 & 0 & 0 & 0 & 0 & 0 \\
\end{pmatrix}\otimes\begin{pmatrix}
 1 & 0 & 1 & 0 & 1 & 0 \\
 1 & 0 & 1 & 0 & 1 & 0 \\
 1 & 0 & 1 & 0 & 1 & 0 \\
 1 & 0 & 1 & 0 & 1 & 0 \\
 1 & 0 & 1 & 0 & 1 & 0 \\
 1 & 0 & 1 & 0 & 1 & 0 \\
\end{pmatrix}\otimes\begin{pmatrix}
 0 & 0 \\
 \bar1 & 0 \\
 1 & 0 \\
 0 & 0 \\
 0 & 0 \\
 1 & 0 \\
\end{pmatrix}\\+
\begin{pmatrix}
 1 & 0 & 1 & \bar1 & 0 & \bar1 \\
 0 & 0 & 0 & 0 & 0 & 0 \\
\end{pmatrix}\otimes\begin{pmatrix}
 0 & 0 & 0 & 0 & 0 & 0 \\
 1 & 0 & 0 & 1 & 1 & 0 \\
 1 & 0 & 0 & 1 & 1 & 0 \\
 0 & 0 & 0 & 0 & 0 & 0 \\
 1 & 0 & 0 & 1 & 1 & 0 \\
 0 & 0 & 0 & 0 & 0 & 0 \\
\end{pmatrix}\otimes\begin{pmatrix}
 0 & 0 \\
 0 & 0 \\
 \bar1 & 0 \\
 1 & 0 \\
 0 & 0 \\
 0 & 0 \\
\end{pmatrix}&&+
\begin{pmatrix}
 \bar1 & 0 & 0 & 1 & 0 & 0 \\
 0 & 0 & \bar1 & 0 & 0 & 1 \\
\end{pmatrix}\otimes\begin{pmatrix}
 0 & 0 & 0 & 1 & 0 & 0 \\
 0 & 0 & 0 & 1 & 0 & 0 \\
 0 & 1 & 0 & 0 & 0 & 1 \\
 0 & 1 & 0 & 0 & 0 & 1 \\
 0 & 0 & 0 & 0 & 0 & 0 \\
 0 & 1 & 0 & 1 & 0 & 1 \\
\end{pmatrix}\otimes\begin{pmatrix}
 0 & 0 \\
 0 & 0 \\
 1 & 0 \\
 0 & 1 \\
 0 & 0 \\
 1 & 1 \\
\end{pmatrix}
\end{alignat*}
\normalsize

Here is one of our schemes of rank 56 for the format $(3,4,6)$. Again, minus signs are
placed above numbers rather than in front of them, e.g., $\bar 2$ means $-2$. 

\footnotesize
\begin{alignat*}1
&\begin{pmatrix}
 0 & 0 & \bar1 & 0 \\
 0 & 0 & 0 & 0 \\
 0 & 0 & 0 & 0 \\
\end{pmatrix}
\otimes
\begin{pmatrix}
 \bar2 & \bar2 & 1 & 0 & \bar2 & \bar2 \\
 0 & 0 & \bar1 & 0 & 0 & 0 \\
 \bar1 & \bar1 & 0 & 0 & \bar1 & \bar1 \\
 1 & 1 & \bar1 & 0 & 1 & 1 \\
\end{pmatrix}
\otimes
\begin{pmatrix}
 1 & 0 & 0 \\
 0 & 0 & 0 \\
 0 & 0 & 0 \\
 0 & 0 & 0 \\
 0 & 0 & 0 \\
 0 & 0 & 0 \\
\end{pmatrix}
+
\begin{pmatrix}
 1 & 0 & \bar1 & 1 \\
 0 & 0 & \bar1 & 0 \\
 1 & 0 & \bar1 & 1 \\
\end{pmatrix}
\otimes
\begin{pmatrix}
 \bar1 & 0 & 1 & 0 & \bar1 & \bar1 \\
 0 & 1 & 0 & 0 & 0 & 0 \\
 0 & 1 & 0 & 0 & 0 & 0 \\
 0 & 0 & 0 & 0 & 0 & 0 \\
\end{pmatrix}
\otimes
\begin{pmatrix}
 0 & 0 & \bar1 \\
 0 & \bar1 & 0 \\
 0 & 0 & 0 \\
 0 & \bar1 & 1 \\
 0 & 1 & 0 \\
 0 & 0 & 0 \\
\end{pmatrix}
\\+&
\begin{pmatrix}
 \bar1 & 0 & 1 & 0 \\
 0 & 0 & 0 & 0 \\
 \bar1 & 0 & 1 & 0 \\
\end{pmatrix}
\otimes
\begin{pmatrix}
 0 & 0 & 1 & 0 & 0 & 0 \\
 0 & 0 & 0 & 0 & 0 & 0 \\
 0 & 0 & 0 & 0 & 0 & 0 \\
 0 & 0 & 0 & 0 & 0 & 0 \\
\end{pmatrix}
\otimes
\begin{pmatrix}
 1 & \bar1 & 1 \\
 \bar1 & 0 & 0 \\
 \bar1 & 1 & 0 \\
 0 & 0 & 0 \\
 0 & 1 & \bar1 \\
 \bar1 & 1 & 0 \\
\end{pmatrix}
+
\begin{pmatrix}
 0 & \bar1 & 0 & 0 \\
 0 & 0 & 0 & 0 \\
 0 & \bar1 & 1 & 0 \\
\end{pmatrix}
\otimes
\begin{pmatrix}
 \bar1 & \bar1 & 1 & 1 & \bar1 & \bar1 \\
 1 & 1 & \bar1 & 1 & 1 & 1 \\
 0 & 0 & 0 & 1 & 0 & 0 \\
 1 & 1 & \bar1 & 0 & 1 & 1 \\
\end{pmatrix}
\otimes
\begin{pmatrix}
 0 & 0 & \bar1 \\
 0 & 0 & 0 \\
 0 & 0 & 0 \\
 \bar1 & 0 & 1 \\
 0 & 0 & 0 \\
 0 & 0 & 0 \\
\end{pmatrix}
\\+&
\begin{pmatrix}
 0 & \bar1 & 1 & 1 \\
 0 & 0 & 0 & 0 \\
 0 & \bar1 & 0 & 1 \\
\end{pmatrix}
\otimes
\begin{pmatrix}
 \bar1 & \bar1 & 0 & 0 & \bar1 & \bar1 \\
 0 & 0 & \bar1 & 0 & 0 & 0 \\
 0 & 0 & \bar1 & 0 & 0 & 0 \\
 1 & 1 & \bar1 & 0 & 1 & 1 \\
\end{pmatrix}
\otimes
\begin{pmatrix}
 1 & 0 & 0 \\
 0 & 0 & 0 \\
 \bar1 & 0 & 1 \\
 0 & 0 & 0 \\
 0 & 0 & 0 \\
 \bar1 & 0 & 1 \\
\end{pmatrix}
+
\begin{pmatrix}
 0 & 0 & 0 & \bar1 \\
 0 & 0 & 0 & 0 \\
 0 & 0 & 0 & 0 \\
\end{pmatrix}
\otimes
\begin{pmatrix}
 0 & 0 & 0 & 0 & 0 & 0 \\
 0 & 0 & 1 & 0 & 0 & 0 \\
 0 & 0 & 0 & 0 & 0 & 0 \\
 0 & 0 & 1 & 0 & 0 & 0 \\
\end{pmatrix}
\otimes
\begin{pmatrix}
 \bar1 & 1 & \bar1 \\
 0 & 1 & 0 \\
 \bar1 & 1 & 0 \\
 1 & 0 & 0 \\
 0 & \bar1 & 1 \\
 0 & 0 & 0 \\
\end{pmatrix}
\\+&
\begin{pmatrix}
 0 & 0 & 0 & 0 \\
 0 & 0 & \bar1 & 0 \\
 0 & 0 & 0 & 0 \\
\end{pmatrix}
\otimes
\begin{pmatrix}
 1 & 0 & \bar1 & 1 & 1 & 1 \\
 0 & \bar1 & 0 & 1 & 0 & 0 \\
 0 & \bar1 & 0 & 1 & 0 & 0 \\
 0 & 0 & 0 & 0 & 0 & 0 \\
\end{pmatrix}
\otimes
\begin{pmatrix}
 0 & 1 & \bar1 \\
 0 & 0 & 0 \\
 0 & 0 & 0 \\
 0 & \bar1 & 1 \\
 0 & 0 & 0 \\
 0 & 0 & 0 \\
\end{pmatrix}
+
\begin{pmatrix}
 0 & 0 & 0 & 0 \\
 1 & \bar1 & 0 & 1 \\
 0 & 0 & 0 & 0 \\
\end{pmatrix}
\otimes
\begin{pmatrix}
 0 & 0 & \bar2 & 0 & 0 & 1 \\
 0 & \bar1 & \bar1 & 0 & \bar1 & 1 \\
 0 & \bar1 & \bar1 & 0 & \bar1 & 1 \\
 0 & 0 & 0 & 0 & 0 & 0 \\
\end{pmatrix}
\otimes
\begin{pmatrix}
 0 & \bar1 & 0 \\
 0 & 1 & 0 \\
 0 & 0 & 0 \\
 0 & 0 & 0 \\
 0 & 0 & 0 \\
 0 & 0 & 0 \\
\end{pmatrix}
\\+&
\begin{pmatrix}
 0 & 0 & 0 & 1 \\
 0 & 0 & 0 & 1 \\
 0 & 0 & 0 & 0 \\
\end{pmatrix}
\otimes
\begin{pmatrix}
 \bar2 & 0 & \bar1 & \bar1 & \bar1 & \bar1 \\
 \bar1 & 0 & \bar1 & \bar1 & \bar1 & 0 \\
 \bar1 & 0 & \bar1 & \bar1 & \bar1 & 0 \\
 1 & 0 & 0 & 0 & 0 & 1 \\
\end{pmatrix}
\otimes
\begin{pmatrix}
 0 & 0 & 0 \\
 0 & 0 & 0 \\
 0 & 1 & 0 \\
 0 & 0 & 0 \\
 0 & 0 & 0 \\
 0 & 1 & 0 \\
\end{pmatrix}
+
\begin{pmatrix}
 0 & 1 & 0 & 0 \\
 0 & 1 & \bar1 & 0 \\
 0 & 0 & 0 & 0 \\
\end{pmatrix}
\otimes
\begin{pmatrix}
 \bar1 & \bar1 & 2 & 0 & \bar1 & \bar2 \\
 1 & 1 & 0 & 0 & 1 & 0 \\
 0 & 0 & 1 & 0 & 0 & \bar1 \\
 1 & 1 & \bar1 & 0 & 1 & 1 \\
\end{pmatrix}
\otimes
\begin{pmatrix}
 2 & \bar1 & 0 \\
 0 & 0 & 0 \\
 0 & 0 & 0 \\
 0 & 0 & 0 \\
 \bar1 & 1 & 0 \\
 \bar1 & 1 & 0 \\
\end{pmatrix}
\\+&
\begin{pmatrix}
 0 & 0 & 0 & 0 \\
 \bar1 & 1 & 0 & \bar1 \\
 0 & 0 & 0 & 0 \\
\end{pmatrix}
\otimes
\begin{pmatrix}
 1 & 0 & \bar2 & 1 & 1 & 1 \\
 0 & \bar1 & \bar1 & 0 & \bar1 & 1 \\
 0 & \bar1 & \bar1 & 0 & \bar1 & 1 \\
 0 & 0 & 0 & 0 & 0 & 0 \\
\end{pmatrix}
\otimes
\begin{pmatrix}
 0 & 0 & \bar1 \\
 0 & 1 & 0 \\
 0 & 0 & 0 \\
 0 & 0 & 0 \\
 0 & \bar1 & 1 \\
 0 & 0 & 0 \\
\end{pmatrix}
+
\begin{pmatrix}
 1 & 0 & \bar1 & 0 \\
 1 & 0 & \bar1 & 0 \\
 1 & 0 & \bar1 & 0 \\
\end{pmatrix}
\otimes
\begin{pmatrix}
 0 & 0 & 1 & 0 & 1 & \bar1 \\
 0 & 0 & 0 & 0 & 0 & 0 \\
 0 & 0 & 0 & 0 & 0 & 0 \\
 0 & 0 & \bar1 & 0 & \bar1 & 1 \\
\end{pmatrix}
\otimes
\begin{pmatrix}
 0 & 0 & 0 \\
 0 & 1 & 0 \\
 0 & 1 & 0 \\
 0 & 0 & 0 \\
 0 & 0 & 0 \\
 0 & 0 & 0 \\
\end{pmatrix}
\\+&
\begin{pmatrix}
 0 & 1 & \bar1 & 0 \\
 0 & 1 & \bar1 & 0 \\
 0 & 0 & 0 & 0 \\
\end{pmatrix}
\otimes
\begin{pmatrix}
 \bar1 & 0 & 1 & 0 & \bar1 & \bar2 \\
 0 & 0 & \bar1 & 0 & 0 & 0 \\
 0 & 0 & 1 & 0 & 0 & \bar1 \\
 1 & 0 & \bar1 & 0 & 1 & 1 \\
\end{pmatrix}
\otimes
\begin{pmatrix}
 \bar2 & 0 & 0 \\
 0 & 0 & 0 \\
 0 & 0 & 0 \\
 0 & 0 & 0 \\
 1 & 0 & 0 \\
 1 & 0 & 0 \\
\end{pmatrix}
+
\begin{pmatrix}
 0 & \bar1 & 0 & 0 \\
 1 & \bar1 & 0 & 1 \\
 0 & \bar1 & 0 & 0 \\
\end{pmatrix}
\otimes
\begin{pmatrix}
 0 & 0 & 0 & 0 & 0 & 0 \\
 0 & 0 & \bar1 & 0 & \bar1 & 1 \\
 0 & 0 & 0 & 0 & 0 & 0 \\
 0 & 0 & 0 & 0 & 0 & 0 \\
\end{pmatrix}
\otimes
\begin{pmatrix}
 0 & 0 & \bar1 \\
 0 & 0 & 0 \\
 0 & 0 & 0 \\
 0 & 0 & 0 \\
 0 & 0 & 1 \\
 0 & 0 & 0 \\
\end{pmatrix}
\\+&
\begin{pmatrix}
 0 & 0 & 0 & 0 \\
 0 & 0 & 0 & 0 \\
 \bar1 & 0 & 1 & 0 \\
\end{pmatrix}
\otimes
\begin{pmatrix}
 0 & \bar1 & 0 & 0 & 0 & 0 \\
 0 & 0 & 0 & 0 & 0 & 0 \\
 0 & 0 & 0 & 0 & 0 & 0 \\
 0 & 1 & 0 & 0 & 0 & 0 \\
\end{pmatrix}
\otimes
\begin{pmatrix}
 1 & \bar1 & 1 \\
 \bar1 & \bar1 & 1 \\
 \bar1 & 0 & 1 \\
 0 & 0 & 0 \\
 0 & 1 & \bar1 \\
 \bar1 & 1 & 0 \\
\end{pmatrix}
+
\begin{pmatrix}
 0 & 0 & 0 & 0 \\
 0 & 1 & \bar1 & \bar1 \\
 0 & 1 & \bar1 & \bar1 \\
\end{pmatrix}
\otimes
\begin{pmatrix}
 0 & \bar1 & 1 & 0 & 0 & \bar1 \\
 0 & 0 & 0 & 0 & 0 & 0 \\
 1 & 0 & 0 & 1 & 1 & 0 \\
 0 & 1 & 0 & 0 & 0 & 0 \\
\end{pmatrix}
\otimes
\begin{pmatrix}
 0 & 0 & \bar2 \\
 0 & 0 & 0 \\
 0 & 0 & 0 \\
 0 & 0 & 0 \\
 0 & 0 & 1 \\
 0 & 0 & 1 \\
\end{pmatrix}
\\+&
\begin{pmatrix}
 \bar1 & 0 & 1 & \bar1 \\
 0 & 0 & 0 & 0 \\
 0 & 0 & 0 & 0 \\
\end{pmatrix}
\otimes
\begin{pmatrix}
 \bar1 & 0 & 2 & 0 & 0 & \bar2 \\
 0 & 0 & 1 & 0 & 0 & 0 \\
 0 & 1 & 1 & 0 & 1 & \bar1 \\
 0 & \bar1 & 0 & 0 & \bar1 & 1 \\
\end{pmatrix}
\otimes
\begin{pmatrix}
 1 & \bar1 & 1 \\
 0 & \bar1 & 0 \\
 0 & 0 & 0 \\
 0 & 0 & 0 \\
 0 & 1 & \bar1 \\
 \bar1 & 1 & 0 \\
\end{pmatrix}
+
\begin{pmatrix}
 \bar1 & 0 & 1 & \bar1 \\
 0 & 0 & 0 & 0 \\
 \bar1 & 0 & 0 & \bar1 \\
\end{pmatrix}
\otimes
\begin{pmatrix}
 0 & 1 & 0 & 0 & 0 & 0 \\
 0 & 1 & 0 & 0 & 0 & 0 \\
 0 & 1 & 0 & 0 & 0 & 0 \\
 0 & 0 & 0 & 0 & 0 & 0 \\
\end{pmatrix}
\otimes
\begin{pmatrix}
 0 & 0 & \bar1 \\
 1 & 0 & \bar1 \\
 0 & 0 & 0 \\
 0 & 0 & 0 \\
 \bar1 & 0 & 1 \\
 0 & 0 & 0 \\
\end{pmatrix}
\\+&
\begin{pmatrix}
 0 & 0 & 0 & 0 \\
 0 & 1 & \bar1 & \bar1 \\
 0 & 0 & 0 & \bar1 \\
\end{pmatrix}
\otimes
\begin{pmatrix}
 0 & 0 & 0 & 1 & 0 & 0 \\
 1 & 0 & 0 & 1 & 1 & 0 \\
 0 & 0 & 0 & 0 & 0 & 0 \\
 1 & 0 & 0 & 0 & 1 & 0 \\
\end{pmatrix}
\otimes
\begin{pmatrix}
 0 & 0 & \bar2 \\
 0 & 0 & 0 \\
 0 & 0 & 0 \\
 0 & 1 & 0 \\
 0 & 0 & 1 \\
 0 & 0 & 1 \\
\end{pmatrix}
+
\begin{pmatrix}
 \bar1 & 0 & 1 & \bar1 \\
 \bar1 & 0 & 1 & \bar1 \\
 \bar1 & 0 & 1 & \bar1 \\
\end{pmatrix}
\otimes
\begin{pmatrix}
 \bar1 & 0 & 2 & 0 & 0 & \bar2 \\
 0 & 1 & 0 & 0 & 0 & 0 \\
 0 & 1 & 1 & 0 & 1 & \bar1 \\
 0 & 0 & \bar1 & 0 & \bar1 & 1 \\
\end{pmatrix}
\otimes
\begin{pmatrix}
 0 & 0 & 0 \\
 0 & \bar1 & 0 \\
 0 & 0 & 0 \\
 0 & \bar1 & 0 \\
 0 & 1 & 0 \\
 0 & 0 & 0 \\
\end{pmatrix}
\\+&
\begin{pmatrix}
 \bar1 & 0 & 1 & 0 \\
 \bar1 & 1 & 0 & \bar1 \\
 \bar1 & 0 & 1 & 0 \\
\end{pmatrix}
\otimes
\begin{pmatrix}
 0 & 0 & 0 & 0 & 1 & \bar1 \\
 0 & 0 & 0 & 0 & 0 & 0 \\
 0 & 0 & 0 & 0 & 0 & 0 \\
 0 & 0 & \bar1 & 0 & \bar1 & 1 \\
\end{pmatrix}
\otimes
\begin{pmatrix}
 0 & \bar1 & 1 \\
 0 & 0 & 0 \\
 0 & 1 & 0 \\
 0 & 0 & 0 \\
 0 & 1 & \bar1 \\
 0 & 1 & 0 \\
\end{pmatrix}
+
\begin{pmatrix}
 \bar1 & 1 & 0 & \bar1 \\
 0 & 0 & 0 & 0 \\
 0 & 0 & 0 & 0 \\
\end{pmatrix}
\otimes
\begin{pmatrix}
 0 & \bar2 & 0 & 0 & \bar1 & 1 \\
 0 & 0 & 0 & 0 & 0 & 0 \\
 0 & \bar1 & \bar1 & 0 & \bar1 & 1 \\
 0 & 1 & 0 & 0 & 0 & 0 \\
\end{pmatrix}
\otimes
\begin{pmatrix}
 \bar1 & 0 & 0 \\
 1 & 0 & 0 \\
 0 & 0 & 0 \\
 0 & 0 & 0 \\
 0 & 0 & 0 \\
 0 & 0 & 0 \\
\end{pmatrix}
\\+&
\begin{pmatrix}
 1 & 0 & \bar1 & 1 \\
 1 & 1 & \bar2 & 0 \\
 1 & 0 & \bar1 & 0 \\
\end{pmatrix}
\otimes
\begin{pmatrix}
 0 & 0 & 0 & 0 & 1 & 0 \\
 0 & 0 & 0 & 0 & 0 & 0 \\
 0 & 0 & 0 & 0 & 0 & 0 \\
 0 & 0 & \bar1 & 0 & \bar1 & 1 \\
\end{pmatrix}
\otimes
\begin{pmatrix}
 0 & \bar1 & 0 \\
 0 & \bar1 & 0 \\
 0 & 0 & 0 \\
 0 & 0 & 0 \\
 0 & 1 & 0 \\
 0 & 1 & 0 \\
\end{pmatrix}
+
\begin{pmatrix}
 0 & \bar1 & 0 & 0 \\
 0 & 0 & 0 & 0 \\
 0 & 0 & 0 & 0 \\
\end{pmatrix}
\otimes
\begin{pmatrix}
 \bar1 & 0 & 2 & 0 & 0 & \bar2 \\
 1 & 2 & 0 & 0 & 2 & 0 \\
 0 & 1 & 1 & 0 & 1 & \bar1 \\
 1 & 1 & \bar1 & 0 & 1 & 1 \\
\end{pmatrix}
\otimes
\begin{pmatrix}
 1 & \bar1 & 1 \\
 0 & 0 & 0 \\
 0 & 0 & 0 \\
 1 & 0 & \bar1 \\
 \bar1 & 1 & 0 \\
 \bar1 & 1 & 0 \\
\end{pmatrix}
\\+&
\begin{pmatrix}
 1 & 0 & \bar1 & 1 \\
 0 & 0 & 0 & 0 \\
 0 & 0 & 0 & 0 \\
\end{pmatrix}
\otimes
\begin{pmatrix}
 0 & 0 & 0 & 0 & 0 & 0 \\
 0 & 1 & 0 & 0 & 0 & 0 \\
 0 & 0 & 0 & 0 & 0 & 0 \\
 0 & 1 & 0 & 0 & 0 & 0 \\
\end{pmatrix}
\otimes
\begin{pmatrix}
 0 & 0 & 0 \\
 1 & 0 & \bar1 \\
 0 & 0 & 0 \\
 1 & 0 & \bar1 \\
 \bar1 & 0 & 1 \\
 0 & 0 & 0 \\
\end{pmatrix}
+
\begin{pmatrix}
 \bar1 & 0 & 0 & \bar1 \\
 0 & 0 & 0 & 0 \\
 \bar1 & 0 & 0 & \bar1 \\
\end{pmatrix}
\otimes
\begin{pmatrix}
 0 & 0 & \bar1 & 0 & \bar1 & 1 \\
 0 & 0 & 0 & 0 & 0 & 0 \\
 0 & 0 & \bar1 & 0 & \bar1 & 1 \\
 0 & 0 & 0 & 0 & 0 & 0 \\
\end{pmatrix}
\otimes
\begin{pmatrix}
 0 & 0 & 0 \\
 \bar1 & 0 & 0 \\
 0 & 0 & 0 \\
 0 & 0 & 0 \\
 1 & 0 & 0 \\
 0 & 0 & 0 \\
\end{pmatrix}
\\+&
\begin{pmatrix}
 0 & \bar1 & 0 & 0 \\
 \bar1 & 0 & 0 & \bar1 \\
 0 & 0 & 0 & 0 \\
\end{pmatrix}
\otimes
\begin{pmatrix}
 0 & \bar1 & \bar1 & 0 & \bar1 & 1 \\
 0 & \bar1 & \bar1 & 0 & \bar1 & 1 \\
 0 & \bar1 & \bar1 & 0 & \bar1 & 1 \\
 0 & 0 & 0 & 0 & 0 & 0 \\
\end{pmatrix}
\otimes
\begin{pmatrix}
 0 & \bar1 & 2 \\
 0 & 0 & 0 \\
 0 & 0 & 0 \\
 1 & 0 & \bar1 \\
 0 & 1 & \bar1 \\
 \bar1 & 1 & 0 \\
\end{pmatrix}
+
\begin{pmatrix}
 0 & 1 & \bar1 & \bar1 \\
 0 & 0 & 0 & 0 \\
 0 & 1 & \bar1 & 0 \\
\end{pmatrix}
\otimes
\begin{pmatrix}
 0 & 0 & 0 & \bar1 & 0 & 0 \\
 0 & 0 & 1 & 0 & 0 & 0 \\
 0 & 0 & 0 & 0 & 0 & 0 \\
 0 & 0 & 1 & 1 & 0 & 0 \\
\end{pmatrix}
\otimes
\begin{pmatrix}
 0 & 0 & 0 \\
 0 & 0 & 0 \\
 0 & 0 & 1 \\
 \bar1 & 0 & 0 \\
 0 & 0 & 0 \\
 0 & 0 & 1 \\
\end{pmatrix}
\\+&
\begin{pmatrix}
 0 & 0 & 0 & 0 \\
 0 & 1 & \bar1 & \bar1 \\
 0 & 0 & 0 & 0 \\
\end{pmatrix}
\otimes
\begin{pmatrix}
 0 & \bar1 & 1 & 0 & 0 & \bar1 \\
 0 & 0 & 0 & 0 & 0 & 0 \\
 0 & 0 & 0 & 0 & 0 & 0 \\
 0 & 1 & \bar1 & 0 & 0 & 1 \\
\end{pmatrix}
\otimes
\begin{pmatrix}
 0 & 2 & \bar2 \\
 0 & 0 & 0 \\
 0 & \bar1 & 0 \\
 0 & 0 & 0 \\
 0 & \bar1 & 1 \\
 0 & \bar2 & 1 \\
\end{pmatrix}
+
\begin{pmatrix}
 0 & 0 & 0 & 0 \\
 0 & \bar1 & 1 & 1 \\
 \bar1 & 0 & 1 & 0 \\
\end{pmatrix}
\otimes
\begin{pmatrix}
 0 & \bar1 & 1 & 0 & 0 & \bar1 \\
 0 & 0 & 0 & 0 & 0 & 0 \\
 0 & 0 & 0 & 0 & 0 & 0 \\
 0 & 1 & 0 & 0 & 0 & 0 \\
\end{pmatrix}
\otimes
\begin{pmatrix}
 0 & 1 & \bar2 \\
 0 & 1 & 0 \\
 0 & 0 & 0 \\
 0 & 0 & 0 \\
 0 & \bar1 & 1 \\
 0 & \bar1 & 1 \\
\end{pmatrix}
\\+&
\begin{pmatrix}
 0 & 0 & 0 & 0 \\
 0 & 0 & 0 & 1 \\
 0 & 0 & 0 & 1 \\
\end{pmatrix}
\otimes
\begin{pmatrix}
 0 & 0 & 0 & 0 & 0 & 0 \\
 1 & 0 & 0 & 1 & 1 & 0 \\
 0 & 0 & 0 & 0 & 0 & 0 \\
 1 & 0 & 0 & 1 & 1 & 0 \\
\end{pmatrix}
\otimes
\begin{pmatrix}
 0 & 0 & 0 \\
 0 & 0 & 0 \\
 0 & 0 & 0 \\
 0 & 1 & 0 \\
 0 & 0 & 0 \\
 0 & 0 & 0 \\
\end{pmatrix}
+
\begin{pmatrix}
 0 & 1 & \bar1 & \bar1 \\
 0 & 0 & 0 & 0 \\
 1 & 0 & \bar1 & 0 \\
\end{pmatrix}
\otimes
\begin{pmatrix}
 0 & \bar1 & 1 & 0 & 0 & 0 \\
 0 & 0 & 0 & 0 & 0 & 0 \\
 0 & 0 & 0 & 0 & 0 & 0 \\
 0 & 1 & 0 & 0 & 0 & 0 \\
\end{pmatrix}
\otimes
\begin{pmatrix}
 1 & 0 & 0 \\
 \bar1 & 0 & 0 \\
 \bar1 & 0 & 1 \\
 0 & 0 & 0 \\
 0 & 0 & 0 \\
 \bar1 & 0 & 1 \\
\end{pmatrix}
\\+&
\begin{pmatrix}
 0 & 0 & 0 & 0 \\
 0 & \bar1 & 1 & 0 \\
 0 & 0 & 0 & 0 \\
\end{pmatrix}
\otimes
\begin{pmatrix}
 \bar1 & 0 & 2 & 0 & 0 & \bar2 \\
 1 & 1 & 0 & 0 & 1 & 0 \\
 0 & 1 & 1 & 0 & 1 & \bar1 \\
 1 & 0 & \bar1 & 0 & 0 & 1 \\
\end{pmatrix}
\otimes
\begin{pmatrix}
 2 & \bar2 & 0 \\
 0 & 0 & 0 \\
 0 & 0 & 0 \\
 0 & 1 & 0 \\
 \bar1 & 1 & 0 \\
 \bar1 & 1 & 0 \\
\end{pmatrix}
+
\begin{pmatrix}
 \bar1 & 0 & 1 & \bar1 \\
 \bar1 & 1 & 0 & \bar1 \\
 \bar1 & 0 & 1 & \bar1 \\
\end{pmatrix}
\otimes
\begin{pmatrix}
 \bar1 & 0 & 2 & \bar1 & 0 & \bar2 \\
 0 & 1 & 0 & 0 & 0 & 0 \\
 0 & 1 & 1 & 0 & 1 & \bar1 \\
 0 & 0 & \bar1 & 0 & \bar1 & 1 \\
\end{pmatrix}
\otimes
\begin{pmatrix}
 0 & 0 & \bar1 \\
 0 & 1 & 0 \\
 0 & 0 & 0 \\
 0 & 1 & 0 \\
 0 & \bar1 & 1 \\
 0 & 0 & 0 \\
\end{pmatrix}
\\+&
\begin{pmatrix}
 1 & 0 & \bar1 & 0 \\
 0 & 0 & 0 & 0 \\
 0 & 0 & 0 & 0 \\
\end{pmatrix}
\otimes
\begin{pmatrix}
 0 & 0 & 1 & 0 & 1 & \bar1 \\
 0 & 0 & 1 & 0 & 0 & 0 \\
 0 & 0 & 0 & 0 & 0 & 0 \\
 0 & 0 & 0 & 0 & \bar1 & 1 \\
\end{pmatrix}
\otimes
\begin{pmatrix}
 1 & \bar1 & 1 \\
 0 & \bar1 & 0 \\
 0 & 0 & 0 \\
 0 & 0 & 0 \\
 0 & 1 & \bar1 \\
 \bar1 & 1 & 0 \\
\end{pmatrix}
+
\begin{pmatrix}
 0 & 0 & 0 & 0 \\
 0 & 0 & 0 & 0 \\
 0 & 1 & 0 & \bar1 \\
\end{pmatrix}
\otimes
\begin{pmatrix}
 \bar1 & \bar1 & 1 & 0 & \bar1 & \bar1 \\
 0 & 0 & 0 & 0 & 0 & 0 \\
 0 & 0 & 0 & 0 & 0 & 0 \\
 1 & 1 & \bar1 & 0 & 1 & 1 \\
\end{pmatrix}
\otimes
\begin{pmatrix}
 1 & 1 & \bar1 \\
 0 & 0 & 0 \\
 \bar1 & \bar1 & 1 \\
 0 & 0 & 0 \\
 0 & 0 & 0 \\
 \bar1 & \bar1 & 1 \\
\end{pmatrix}
\\+&
\begin{pmatrix}
 1 & 0 & \bar1 & 1 \\
 0 & 0 & 0 & 0 \\
 1 & 0 & \bar1 & 1 \\
\end{pmatrix}
\otimes
\begin{pmatrix}
 \bar1 & 0 & 1 & \bar1 & \bar1 & \bar1 \\
 0 & 1 & 0 & 0 & 0 & 0 \\
 0 & 1 & 0 & 0 & 0 & 0 \\
 0 & 0 & 0 & 0 & 0 & 0 \\
\end{pmatrix}
\otimes
\begin{pmatrix}
 0 & 0 & 0 \\
 0 & 1 & \bar1 \\
 0 & 0 & 0 \\
 0 & 1 & \bar1 \\
 0 & \bar1 & 1 \\
 0 & 0 & 0 \\
\end{pmatrix}
+
\begin{pmatrix}
 0 & 1 & \bar1 & 0 \\
 0 & 0 & 0 & 0 \\
 0 & 1 & \bar1 & 0 \\
\end{pmatrix}
\otimes
\begin{pmatrix}
 0 & 0 & 0 & \bar2 & 0 & 0 \\
 0 & 0 & 1 & 0 & 0 & 0 \\
 0 & 0 & 0 & \bar1 & 0 & 0 \\
 0 & 0 & 1 & 1 & 0 & 0 \\
\end{pmatrix}
\otimes
\begin{pmatrix}
 0 & 0 & 0 \\
 0 & 0 & 0 \\
 0 & 0 & 0 \\
 1 & 0 & 0 \\
 0 & 0 & 0 \\
 0 & 0 & 0 \\
\end{pmatrix}
\\+&
\begin{pmatrix}
 0 & 1 & \bar1 & \bar1 \\
 0 & 0 & 0 & 0 \\
 0 & 0 & 0 & 0 \\
\end{pmatrix}
\otimes
\begin{pmatrix}
 \bar1 & 0 & 0 & 0 & \bar1 & \bar1 \\
 0 & 0 & \bar1 & 0 & 0 & 0 \\
 0 & 0 & 0 & 0 & 0 & 0 \\
 1 & 0 & \bar1 & 0 & 1 & 1 \\
\end{pmatrix}
\otimes
\begin{pmatrix}
 2 & 0 & 0 \\
 0 & 0 & 0 \\
 \bar1 & 0 & 1 \\
 0 & 0 & 0 \\
 \bar1 & 0 & 0 \\
 \bar2 & 0 & 1 \\
\end{pmatrix}
+
\begin{pmatrix}
 \bar1 & 0 & 1 & \bar1 \\
 \bar1 & 0 & 0 & \bar1 \\
 0 & 0 & 0 & 0 \\
\end{pmatrix}
\otimes
\begin{pmatrix}
 1 & 1 & \bar1 & 0 & 1 & 1 \\
 0 & 0 & 0 & 0 & 0 & 0 \\
 0 & 0 & 0 & 0 & 0 & 0 \\
 0 & 0 & 0 & 0 & 0 & 0 \\
\end{pmatrix}
\otimes
\begin{pmatrix}
 0 & \bar1 & 2 \\
 0 & \bar1 & 0 \\
 0 & 0 & 0 \\
 1 & 0 & \bar1 \\
 0 & 1 & \bar1 \\
 \bar1 & 1 & 0 \\
\end{pmatrix}
\\+&
\begin{pmatrix}
 0 & 1 & 0 & 0 \\
 0 & 1 & 0 & 0 \\
 0 & 0 & 0 & 0 \\
\end{pmatrix}
\otimes
\begin{pmatrix}
 0 & 0 & \bar1 & 0 & 0 & 1 \\
 0 & 0 & \bar1 & 0 & 0 & 1 \\
 0 & 0 & \bar1 & 0 & 0 & 1 \\
 0 & 0 & 0 & 0 & 0 & 0 \\
\end{pmatrix}
\otimes
\begin{pmatrix}
 0 & \bar1 & 0 \\
 0 & 0 & 0 \\
 0 & 0 & 0 \\
 0 & 0 & 0 \\
 0 & 1 & 0 \\
 0 & 1 & 0 \\
\end{pmatrix}
+
\begin{pmatrix}
 1 & 0 & \bar1 & 1 \\
 0 & 0 & 0 & 0 \\
 0 & 1 & \bar1 & 0 \\
\end{pmatrix}
\otimes
\begin{pmatrix}
 \bar1 & \bar1 & 1 & \bar1 & \bar1 & \bar1 \\
 0 & 1 & 0 & 0 & 0 & 0 \\
 0 & 0 & 0 & 0 & 0 & 0 \\
 0 & 1 & 0 & 0 & 0 & 0 \\
\end{pmatrix}
\otimes
\begin{pmatrix}
 0 & 0 & \bar1 \\
 0 & 0 & 1 \\
 0 & 0 & 0 \\
 \bar1 & 0 & 1 \\
 0 & 0 & 0 \\
 0 & 0 & 0 \\
\end{pmatrix}
\\+&
\begin{pmatrix}
 0 & 0 & 0 & 0 \\
 0 & 0 & 0 & 0 \\
 1 & \bar1 & 0 & 1 \\
\end{pmatrix}
\otimes
\begin{pmatrix}
 \bar1 & \bar1 & 1 & \bar1 & \bar1 & \bar1 \\
 0 & 0 & 0 & 0 & 0 & 0 \\
 0 & 0 & 0 & 0 & 0 & 0 \\
 0 & 1 & 0 & 0 & 0 & 0 \\
\end{pmatrix}
\otimes
\begin{pmatrix}
 0 & 0 & \bar1 \\
 0 & 0 & 1 \\
 0 & 0 & 0 \\
 0 & 0 & 0 \\
 0 & 0 & 0 \\
 0 & 0 & 0 \\
\end{pmatrix}
+
\begin{pmatrix}
 0 & 1 & 0 & \bar1 \\
 0 & 0 & 0 & 0 \\
 0 & 1 & 0 & \bar1 \\
\end{pmatrix}
\otimes
\begin{pmatrix}
 0 & 0 & 1 & 0 & 0 & 0 \\
 0 & 0 & 1 & 0 & 0 & 0 \\
 0 & 0 & 1 & 0 & 0 & 0 \\
 0 & 0 & 0 & 0 & 0 & 0 \\
\end{pmatrix}
\otimes
\begin{pmatrix}
 \bar1 & 0 & 0 \\
 0 & 0 & 0 \\
 1 & 0 & 0 \\
 0 & 0 & 0 \\
 0 & 0 & 0 \\
 1 & 0 & 0 \\
\end{pmatrix}
\\+&
\begin{pmatrix}
 0 & 0 & 0 & 0 \\
 0 & \bar1 & 0 & 1 \\
 0 & \bar1 & 0 & 1 \\
\end{pmatrix}
\otimes
\begin{pmatrix}
 1 & 0 & 0 & 1 & 1 & 0 \\
 1 & 0 & 0 & 1 & 1 & 0 \\
 1 & 0 & 0 & 1 & 1 & 0 \\
 0 & 0 & 0 & 0 & 0 & 0 \\
\end{pmatrix}
\otimes
\begin{pmatrix}
 0 & \bar1 & 0 \\
 0 & 0 & 0 \\
 0 & 1 & 0 \\
 0 & 0 & 0 \\
 0 & 0 & 0 \\
 0 & 1 & 0 \\
\end{pmatrix}
+
\begin{pmatrix}
 0 & \bar1 & 0 & 0 \\
 0 & 0 & 0 & 0 \\
 \bar1 & 0 & 0 & \bar1 \\
\end{pmatrix}
\otimes
\begin{pmatrix}
 0 & \bar1 & \bar1 & 0 & \bar1 & 1 \\
 0 & \bar1 & 0 & 0 & 0 & 0 \\
 0 & \bar1 & \bar1 & 0 & \bar1 & 1 \\
 0 & 0 & 0 & 0 & 0 & 0 \\
\end{pmatrix}
\otimes
\begin{pmatrix}
 0 & 0 & \bar1 \\
 1 & 0 & 0 \\
 0 & 0 & 0 \\
 0 & 0 & 0 \\
 \bar1 & 0 & 1 \\
 0 & 0 & 0 \\
\end{pmatrix}
\\+&
\begin{pmatrix}
 0 & 0 & 0 & 0 \\
 1 & 0 & 0 & 1 \\
 0 & 0 & 0 & 0 \\
\end{pmatrix}
\otimes
\begin{pmatrix}
 1 & 0 & \bar2 & 0 & 0 & 2 \\
 0 & \bar1 & \bar1 & 0 & \bar1 & 1 \\
 0 & \bar1 & \bar1 & 0 & \bar1 & 1 \\
 0 & 0 & 0 & 0 & 0 & 0 \\
\end{pmatrix}
\otimes
\begin{pmatrix}
 0 & 0 & 2 \\
 0 & 0 & 0 \\
 0 & 0 & 0 \\
 1 & 0 & \bar1 \\
 0 & 0 & \bar1 \\
 \bar1 & 1 & 0 \\
\end{pmatrix}
+
\begin{pmatrix}
 0 & 0 & 0 & 0 \\
 0 & 0 & 0 & 0 \\
 0 & 1 & \bar1 & 0 \\
\end{pmatrix}
\otimes
\begin{pmatrix}
 0 & 0 & 0 & 1 & 0 & 0 \\
 1 & 0 & \bar1 & 1 & 1 & 1 \\
 0 & 0 & 0 & 0 & 0 & 0 \\
 1 & 0 & \bar1 & 0 & 1 & 1 \\
\end{pmatrix}
\otimes
\begin{pmatrix}
 0 & 0 & \bar2 \\
 0 & 0 & 0 \\
 0 & 0 & 0 \\
 \bar1 & 0 & 1 \\
 0 & 0 & 1 \\
 0 & 0 & 1 \\
\end{pmatrix}
\\+&
\begin{pmatrix}
 0 & 0 & 0 & 0 \\
 0 & \bar1 & 1 & 1 \\
 0 & \bar1 & 0 & 1 \\
\end{pmatrix}
\otimes
\begin{pmatrix}
 0 & \bar1 & 1 & 1 & 0 & \bar1 \\
 1 & 0 & 0 & 1 & 1 & 0 \\
 1 & 0 & 0 & 1 & 1 & 0 \\
 1 & 1 & \bar1 & 0 & 1 & 1 \\
\end{pmatrix}
\otimes
\begin{pmatrix}
 0 & 1 & \bar2 \\
 0 & 0 & 0 \\
 0 & \bar1 & 0 \\
 0 & 0 & 0 \\
 0 & 0 & 1 \\
 0 & \bar1 & 1 \\
\end{pmatrix}
+
\begin{pmatrix}
 \bar1 & 0 & 1 & \bar1 \\
 0 & \bar1 & 1 & 0 \\
 0 & 0 & 0 & 0 \\
\end{pmatrix}
\otimes
\begin{pmatrix}
 0 & \bar1 & \bar1 & 0 & 0 & 1 \\
 0 & 0 & 0 & 0 & 0 & 0 \\
 0 & \bar1 & \bar1 & 0 & \bar1 & 1 \\
 0 & 1 & 0 & 0 & 0 & 0 \\
\end{pmatrix}
\otimes
\begin{pmatrix}
 2 & \bar1 & 0 \\
 0 & \bar1 & 0 \\
 0 & 0 & 0 \\
 0 & 0 & 0 \\
 \bar1 & 1 & 0 \\
 \bar1 & 1 & 0 \\
\end{pmatrix}
\\+&
\begin{pmatrix}
 0 & 0 & 0 & \bar1 \\
 0 & \bar1 & 1 & 0 \\
 0 & 0 & 0 & 0 \\
\end{pmatrix}
\otimes
\begin{pmatrix}
 \bar1 & 0 & 0 & 0 & 0 & \bar1 \\
 0 & 0 & \bar1 & 0 & 0 & 0 \\
 0 & 0 & 0 & 0 & 0 & 0 \\
 1 & 0 & \bar1 & 0 & 0 & 1 \\
\end{pmatrix}
\otimes
\begin{pmatrix}
 \bar2 & 0 & 0 \\
 0 & 0 & 0 \\
 0 & 1 & 0 \\
 0 & 0 & 0 \\
 1 & 0 & 0 \\
 1 & 1 & 0 \\
\end{pmatrix}
+
\begin{pmatrix}
 0 & 0 & 0 & 0 \\
 0 & 0 & 0 & 0 \\
 0 & 0 & 0 & \bar1 \\
\end{pmatrix}
\otimes
\begin{pmatrix}
 0 & 0 & 0 & \bar1 & 0 & 0 \\
 0 & 0 & 0 & 0 & 0 & 0 \\
 0 & 0 & 0 & 0 & 0 & 0 \\
 0 & 0 & 0 & 1 & 0 & 0 \\
\end{pmatrix}
\otimes
\begin{pmatrix}
 0 & 0 & 0 \\
 0 & 0 & 0 \\
 0 & 0 & 1 \\
 0 & 1 & \bar1 \\
 0 & 0 & 0 \\
 0 & 0 & 1 \\
\end{pmatrix}
\\+&
\begin{pmatrix}
 0 & \bar1 & 1 & 1 \\
 0 & 0 & 0 & 0 \\
 0 & \bar1 & 1 & 1 \\
\end{pmatrix}
\otimes
\begin{pmatrix}
 0 & 0 & 1 & \bar1 & 0 & 0 \\
 0 & 0 & 1 & 0 & 0 & 0 \\
 0 & 0 & 1 & 0 & 0 & 0 \\
 0 & 0 & 1 & 1 & 0 & 0 \\
\end{pmatrix}
\otimes
\begin{pmatrix}
 0 & 0 & 0 \\
 0 & 0 & 0 \\
 0 & 0 & 1 \\
 0 & 0 & 0 \\
 0 & 0 & 0 \\
 0 & 0 & 1 \\
\end{pmatrix}
+
\begin{pmatrix}
 0 & 1 & 0 & 0 \\
 0 & 0 & 1 & 0 \\
 0 & 1 & 0 & 0 \\
\end{pmatrix}
\otimes
\begin{pmatrix}
 0 & 0 & 0 & 1 & 0 & 0 \\
 0 & 0 & 0 & 1 & 0 & 0 \\
 0 & 0 & 0 & 1 & 0 & 0 \\
 0 & 0 & 0 & 0 & 0 & 0 \\
\end{pmatrix}
\otimes
\begin{pmatrix}
 0 & 0 & \bar1 \\
 0 & 0 & 0 \\
 0 & 0 & 0 \\
 0 & 0 & 1 \\
 0 & 0 & 0 \\
 0 & 0 & 0 \\
\end{pmatrix}
\\+&
\begin{pmatrix}
 0 & 0 & 0 & 0 \\
 0 & 0 & 0 & 0 \\
 0 & 0 & 1 & 0 \\
\end{pmatrix}
\otimes
\begin{pmatrix}
 0 & \bar1 & 0 & 1 & 0 & 0 \\
 1 & 0 & \bar1 & 1 & 1 & 1 \\
 1 & 0 & \bar1 & 1 & 1 & 1 \\
 1 & 1 & \bar1 & 0 & 1 & 1 \\
\end{pmatrix}
\otimes
\begin{pmatrix}
 0 & 0 & \bar1 \\
 0 & 0 & 0 \\
 0 & 0 & 0 \\
 0 & 0 & 0 \\
 0 & 0 & 1 \\
 0 & 0 & 1 \\
\end{pmatrix}
+
\begin{pmatrix}
 0 & 0 & 0 & 1 \\
 0 & 0 & 1 & 1 \\
 0 & 0 & 0 & 0 \\
\end{pmatrix}
\otimes
\begin{pmatrix}
 1 & 0 & 1 & 1 & 1 & 0 \\
 1 & 0 & 1 & 1 & 1 & 0 \\
 1 & 0 & 1 & 1 & 1 & 0 \\
 0 & 0 & 0 & 0 & 0 & 0 \\
\end{pmatrix}
\otimes
\begin{pmatrix}
 0 & 0 & 0 \\
 0 & 0 & 0 \\
 0 & 1 & 0 \\
 0 & 0 & 0 \\
 0 & 0 & 0 \\
 0 & 1 & 0 \\
\end{pmatrix}  
\end{alignat*}
\normalsize

Electronic versions of these schemes as well as
schemes for other formats $(n,m,6)$ over $K=\set Z_2$ are
available at \url{https://github.com/jakobmoosbauer/flips.git}.

\end{document}